\documentclass[preprint, amsmath,amssymb, nobibnotes, aps, prb, superscriptaddress, citeautoscript]{revtex4-2}
\usepackage{import}
\usepackage{preamble}
\usepackage{soul}
\usepackage{float}
\usepackage[normalem]{ulem}
\usepackage[version=4]{mhchem}
\usepackage{silence}
\def\QE{\textsc{Quantum ESPRESSO}\,}

\usepackage{booktabs}
\usepackage{rotating}
\usepackage{longtable}
\usepackage{makecell}

\usepackage{jabbrv}
\usepackage{titletoc}
\usepackage{lipsum}

\newcommand{\editorr}[2]{%
  \expandafter\newcommand\csname #1note\endcsname[1]{%
    \textcolor{#2}{(\textbf{#1:} ##1)}}%
  \expandafter\newcommand\csname #1\endcsname[1]{%
    \textcolor{#2}{##1}}%
  \expandafter\newcommand\csname #1cancel\endcsname[1]{%
    \textcolor{#2}{\sout{##1}}}%
  \expandafter\newcommand\csname #1change\endcsname[2]{%
    \textcolor{#2}{\sout{##1} ##2}}%
  \newenvironment{#1text}{\color{#2}}{\color{black}}
}

\editorr{TI}{red}
\editorr{EM}{blue}
\editorr{LB}{red}

\begin{document}

\title{First-principles Hubbard parameters with automated and reproducible workflows}

\author{Lorenzo Bastonero}
\email{lbastone@uni-bremen.de}
\affiliation{U Bremen Excellence Chair, Bremen Center for Computational Materials Science, and MAPEX Center for Materials and Processes, University of Bremen, D-28359 Bremen, Germany}
% ...
\author{Cristiano Malica}
\affiliation{U Bremen Excellence Chair, Bremen Center for Computational Materials Science, and MAPEX Center for Materials and Processes, University of Bremen, D-28359 Bremen, Germany}
% ...
\author{Eric Macke}
\affiliation{U Bremen Excellence Chair, Bremen Center for Computational Materials Science, and MAPEX Center for Materials and Processes, University of Bremen, D-28359 Bremen, Germany}
% ...
\author{Marnik Bercx}
\affiliation{PSI Center for Scientific Computing,
Theory, and Data, and National Centre for Computational Design and Discovery of Novel Materials (MARVEL), 5232 Villigen PSI, Switzerland}
% ...
\author{Sebastiaan P. Huber}
\affiliation{Theory and Simulation of Materials (THEOS), and National Centre for Computational Design and Discovery of Novel Materials (MARVEL), \'Ecole Polytechnique F\'ed\'erale de Lausanne (EPFL), CH-1015 Lausanne, Switzerland}
% ...
\author{Iurii Timrov}
\affiliation{PSI Center for Scientific Computing,
Theory, and Data, and National Centre for Computational Design and Discovery of Novel Materials (MARVEL), 5232 Villigen PSI, Switzerland}
% ...
\author{Nicola Marzari}
\affiliation{U Bremen Excellence Chair, Bremen Center for Computational Materials Science, and MAPEX Center for Materials and Processes, University of Bremen, D-28359 Bremen, Germany}
\affiliation{PSI Center for Scientific Computing,
Theory, and Data, and National Centre for Computational Design and Discovery of Novel Materials (MARVEL), 5232 Villigen PSI, Switzerland}
\affiliation{Theory and Simulation of Materials (THEOS), and National Centre for Computational Design and Discovery of Novel Materials (MARVEL), \'Ecole Polytechnique F\'ed\'erale de Lausanne (EPFL), CH-1015 Lausanne, Switzerland}


\begin{abstract}
We introduce an automated, flexible framework (aiida-hubbard) to self-consistently calculate Hubbard $U$ and $V$ parameters from first-principles. 
By leveraging density-functional perturbation theory, the computation of the Hubbard parameters is efficiently parallelized using multiple concurrent and inexpensive primitive cell calculations. 
Furthermore, the intersite $V$ parameters are defined on-the-fly during the iterative procedure to account for atomic relaxations and diverse coordination environments.
We devise a novel, code-agnostic data structure to store Hubbard related information together with the atomistic structure, to enhance the reproducibility of Hubbard-corrected calculations.
We demonstrate the scalability and reliability of the framework by computing in high-throughput fashion the self-consistent onsite $U$ and intersite $V$ parameters for 115 Li-containing bulk solids with up to 32 atoms in the unit cell. 
Our analysis of the Hubbard parameters calculated reveals a significant correlation of the onsite $U$ values on the oxidation state and coordination environment of the atom on which the Hubbard manifold is centered, while intersite $V$ values exhibit a general decay with increasing interatomic distance.
We find, e.g., that the numerical values of $U$ for the 3d orbitals of Fe and Mn can vary up to 3~eV and 6~eV, respectively; their distribution is characterized by typical shifts of about 0.5~eV and 1.0~eV upon change in oxidation state, or local coordination environment.
For the intersite $V$ a narrower spread is found, with values ranging between 0.2~eV and 1.6~eV when considering transition metal and oxygen interactions.
This framework paves the way for the exploration 
 of redox materials chemistry and high-throughput screening of $d$ and $f$ compounds across diverse research areas, including the discovery and design of novel energy storage materials, as well as other technologically-relevant applications.
\end{abstract}

\flushbottom
\date{\today}
\maketitle
\thispagestyle{empty}

\section*{INTRODUCTION}
Density-functional theory~\cite{Hohenberg1964,Kohn1965} (DFT) has become a workhorse of computational condensed-matter physics, chemistry, and materials science~\cite{Marzari2021}.
Its long-standing success is based on a favorable balance of accuracy and computational efficiency that is achieved by mapping the complex many-body problem of interacting electrons onto an auxiliary system of non-interacting particles moving in an effective potential.
The primary challenge in DFT applications lies in the exchange-correlation (xc) functional, whose exact analytical form is unknown and must therefore be approximated.
Among the numerous xc functionals proposed, the local-density approximation (LDA) and the generalized-gradient approximation (GGA)~\cite{Perdew1992} are the simplest local/semi-local choices, mainly for efficiency reasons.
However, despite their successful applications to a large variety of systems, these functionals have proven much less adequate for the treatment of transition-metal (TM) and rare-earth (RE) compounds.
These issues originate from electron self-interaction errors (SIEs)~\cite{Perdew1981, MoriSanchez2006}, which particularly plague the description of partially occupied and localized $d$ and $f$ states.
Different xc functional flavors have been proposed to cure this flaw: Hubbard corrections to DFT~\cite{Anisimov1991, Liechtenstein1995, Anisimov1997, Dudarev1998, Jr2010, TancogneDejean2020, Lee2020}, meta-GGA functionals, such as SCAN and its variants~\cite{Sun2015, Bartok2019, Furness2020} (as well as SCAN+$U$~\cite{SaiGautam2018, Long2020, Kaczkowski2021, Artrith2022}) and hybrid functionals (e.g., PBE0~\cite{Adamo1999} and HSE06~\cite{Heyd2003, Heyd2006}), to name a few.

In Hubbard-corrected DFT~\cite{Anisimov1991, Liechtenstein1995, Anisimov1997, Dudarev1998}, one or several corrective terms are added to the base DFT xc functional (typically LDA or GGA), whose strength is gauged by the numerical values of the associated \textit{Hubbard parameters}.
The most widespread formulations include the ``on-site'' $U$ terms, which promote localization of electrons on atomic sites; ``inter-site'' $V$ terms stabilizing states between two atoms~\cite{Jr2010}; and Hund's $J$ terms that account for the opposite-spin interactions within a given shell~\cite{Himmetoglu2011, Burgess2023, Burgess2024}.
The unambiguous determination of these parameters can be achieved by recognizing~\cite{Cococcioni2002, Cococcioni2005} that the rotationally invariant formulation of DFT+$U$ provides a natural correction for the spurious curvature of (semi)local functionals, by a removal of a quadratic term and the addition of a linear one.
This heuristic connection, valid in the weak coupling limit between the target Hubbard manifolds and the rest of the electron bath, allows to calculate from first-principles the Hubbard parameters by means of the linear response of the occupation matrices using constrained DFT (LR-cDFT)~\cite{Cococcioni2005}.
Recently, its reformulation in terms of density-functional perturbation theory (DFPT)~\cite{Timrov2018, Timrov2021} boosted its success owing to the replacement of expensive supercells by a computationally less demanding primitive cell with monochromatic perturbations.
Other strategies for computing Hubbard parameters have also been proposed, including Hartree-Fock based methods\cite{Mosey2007, Mosey2008, Andriotis2010, Agapito2015, TancogneDejean2020, Lee2020} and the constrained random phase approximation (cRPA)~\cite{Springer1998, Kotani2000, Aryasetiawan2004, Aryasetiawan2006}. 
Although Hubbard corrections were originally developed to improve the description of strongly correlated materials (typically involving d or f elements), their success primarily derives from the $U$ correction’s ability to enforce piece-wise linearity and remove electronic self-interaction~\cite{Kulik2006}.
This mechanism alleviates the overstabilization of fractional occupations in standard (semi)local functionals -- a problem that arises from the incomplete cancellation between the xc functional and the Hartree term, especially for localized electrons. 
This improvement is further evidenced by the marked qualitative and quantitative enhancements observed in the electronic structure of molecular systems containing a single transition metal atom when DFT+$U$ is applied~\cite{Kulik2006, Kulik2008}.
In this light, the DFT+$U$ correction serves as a self-interaction correction. 

These well-established methods provide frameworks to compute Hubbard parameters, but their outcome strongly depends on the ground state being perturbed. 
In other words, Hubbard parameters computed for an uncorrected DFT ground state (i.e., $U=0$) may differ significantly from the those obtained for a corrected one (i.e., $U>0$).
Obtaining a stable set of Hubbard parameters thus requires a self-consistent cycle in which a new set of Hubbard parameters is evaluated from a corrected DFT$+U(+V)$ ground state that was determined using the Hubbard parameters from the previous step. 
This cycle can also be combined with structural optimizations~\cite{Hsu2009, Cococcioni2019, Timrov2021, Timrov2022a}, thus allowing for mutual consistency between the ionic and electronic DFT$+U(+V)$ ground states.
Hubbard U and V parameters determined using this procedure often lead to significant improvements in electronic structure properties, such as accurate digital changes in oxidation states~\cite{Timrov2022}, even in first-principles molecular dynamics~\cite{Malica2024}, with only a marginal increase in computational cost~\cite{Himmetoglu2013, Mahajan2022, Binci2023, Gebreyesus2023, Haddadi2024, Timrov2023, Bonfa2024, Chang2024, Binci2024}.

While there have been important efforts to automate the procedure for computing Hubbard parameters using LR-cDFT and Hartree-Fock-based methods, to the best of our knowledge no automated workflow exists that allows to determine Hubbard parameters (including the intersite $V$) in a self-consistent fashion.
In a recent high-throughput study on binary oxides~\cite{Moore2024} the authors implemented a workflow in Atomate~\cite{Mathew2017} to compute Hubbard $U$ and Hund's $J$ parameters using the supercell LR-cDFT approach based on finite differences.
MacEnulty and coauthors developed a feature-rich post-processing routine for \textsc{Abinit} that orchestrates the calculation of Hubbard $U$ and Hund's $J$ parameters again relying on LR-cDFT~\cite{MacEnulty2024}. 
A different approach is employed by the ACBN0 functional~\cite{Agapito2015}, implemented in  Octopus~\cite{TancogneDejean2017,TancogneDejean2020} and AFLOW~\cite{Calderon2015,Supka2017}, where the self-consistent  calculation of $U$ and $V$ parameters is performed at runtime during the self-consistent field energy minimization.
This approach is appealing but uses a different assumption for the first-principles calculation of the Hubbard parameters.
Furthermore, the current implementation of intersite interactions~\cite{TancogneDejean2020} in ACBN0 is less flexible with respect to the coordination environment of central atoms due to the use of user-defined (constant) radial cutoffs, which may represent a blocking and error-prone step in the context of high-throughput applications.
Lastly, we also mention the emergence of machine-learning-based techniques for the determination of Hubbard parameters~\cite{Yu2020, Yu2023, Cai2024, Uhrin2025, Das2024}.
While this approach offers a promising path, training of machine learning models requires extensive datasets of Hubbard parameters generated through well-defined and reproducible calculations.
This is crucial, since the effect of Hubbard corrections not only hinges on the numerical values of the Hubbard parameters but also depends on additional boundary conditions such as the choice of Hubbard projectors, the basis set, and the xc functional~\cite{Kulik2008, O’Regan2010, KirchnerHall2021, Wang2016}.

Hence, a robust, flexible, and reliable framework is needed that automates the submission of thousands of jobs, independently handles common errors and also embraces the FAIR (Findable, Accesible, Interoperable, Reuseable) principles of data management~\cite{Wilkinson2016} that ensure a high degree of reproducibility.

In this work, we present \texttt{aiida-hubbard}, a Python package providing an optimized and automated workflow for the structurally self-consistent calculation of Hubbard $U$ and $V$ parameters using the \textsc{HP} code~\cite{Timrov2022} of the \QE distribution~\cite{Giannozzi2009, Giannozzi2017, Giannozzi2020}, which leverages DFPT~\cite{Baroni2001, Timrov2018, Timrov2021}.
The package is devised as a plugin for AiiDA~\cite{Pizzi2016, Huber2020, Uhrin2021}, a well-established scalable computational infrastructure developed to carry out complex computational workflows while facilitating data provenance.
To store the data, we implement \texttt{Hubbard\-Structure\-Data}, a general and flexible data structure in Python that aims at enhancing the reproducibility of Hubbard-corrected DFT calculations.
In \texttt{aiida-hubbard}, the execution of workflows can be fully customized by the user; for instance, it can be specified whether or not the self-consistency cycle shall involve a geometry optimization step (including atomic positions, lattice vectors, or both at the same time).
We demonstrate the scalability and reliability of the package by computing self-consistently the Hubbard $U$ and $V$ parameters of 115 structurally diverse Li-bearing crystalline solids composed of up to five different elements.
Notably, for the successful workflows only in 6\% of the submitted calculations computational errors occurred, all of which were handled successfully without any human intervention.
Importantly, our analysis reveals that both the oxidation state (OS) and the coordination environment of the Hubbard atoms independently affect the numerical values of the self-consistent Hubbard $U$ and $V$ parameters.
%
\\

\noindent 
For the remainder of this section, we briefly summarize the most essential concepts and notations associated with DFT$+U\!+\!V$~\cite{Jr2010, Himmetoglu2013} and DFPT \cite{Timrov2018, Timrov2021} which are used throughout this study.
Fundamentally, the physical justification for Hubbard $U$ and $V$ corrections lies in their capability to mitigate spurious deviations from the piecewise linearity (PWL) of the DFT total energy with respect to fractional addition or removal of charge~\cite{Perdew1982, Cococcioni2002, Cococcioni2005, Kulik2006, MoriSanchez2006, MoriSanchez2009, Zhao2016}, which are related to electron SIEs~\cite{Kulik2006}.
In DFT$+U\!+\!V$, such deviations from PWL are tackled by adding a penalty term $E_{U+V}$ to the Kohn-Sham (KS) DFT energy~\cite{Jr2010}:
\begin{equation}
    E_{\mathrm{DFT}+U\!+\!V} = E_{\mathrm{DFT}} + E_{U+V}~.
    \label{eq:Edft_plus_u}
\end{equation}
$E_{U+V}$ contains two corrections: (i) an onsite term that penalizes the fractional (i.e., non-idempotent) occupation of orbitals centered on atomic sites, and (ii) an intersite term which stabilizes the occupation of states that are linear combinations of atomic orbitals centered on different (usually neighboring) atoms. 
It reads:
\begin{equation}
    E_{U+V} 
    =
    \frac{1}{2} 
    \sum_I \sum_{\sigma m m'} 
    U^I \left( \delta_{m m'} - n^{II \sigma}_{m m'} \right) n^{II \sigma}_{m' m} 
    - \frac{1}{2}
    \sum_{I} \sum_{J (J \ne I)}^* \sum_{\sigma m m'} V^{I J} 
    n^{I J \sigma}_{m m'} n^{J I \sigma}_{m' m}
    ~,
\label{eq:Edftu}
\end{equation}
where $m$ and $m'$ are magnetic quantum numbers associated with the localized manifold being targeted by the correction, $I$ and $J$ are the atomic site indices, while $U^I$ and $V^{I J}$ are the effective onsite and intersite Hubbard parameters, respectively. 
For the second term of Eq.~\eqref{eq:Edftu}, the sum over $J$ is restricted to cover only those neighbors of each atom $I$ for which a $V$ parameter has been specified (as indicated by the star).
For practical calculations, one must define a \textit{Hubbard manifold} to which the $U$ and $V$ corrections are applied. 
Traditionally, onsite manifolds comprise entire valence \textit{d} shells of TM elements and/or \textit{f} shells of lanthanides and actinides.
Other shells, such as the \textit{p}-shells of chalcogenides and halogenides, may also be targeted~\cite{TancogneDejean2020, KirchnerHall2021, Gelin2024}.
Moreover, in some works Hubbard corrections have been applied concurrently to multiple shells localized on the same Hubbard atom~\cite{Jr2010}, or to smaller subsets of the magnetic quantum orbitals of a shell~\cite{Solovyev1996,O’Regan2010, Macke2024}. 
We note that Eq.~\eqref{eq:Edftu} shows the formalism for collinear spin polarization, and refer the reader to Ref.~\cite{Binci2023} for the non-collinear case.
The occupations $n^{I J \sigma}_{m m'}$ are obtained by projecting the KS wavefunctions $\psi^\sigma_{v,\mathbf{k}}(\mathbf{r})$ onto localized orbitals $\phi^{I}_{m}(\mathbf{r})$: 
\begin{equation}
    n^{I J \sigma}_{m m'}
    =
    \sum_{v,\mathbf{k}} 
    f^\sigma_{v,\mathbf{k}}
    \langle \psi^\sigma_{v,\mathbf{k}} | \phi^{J}_{m'} \rangle
    \langle \phi^{I}_{m} | \psi^\sigma_{v,\mathbf{k}} \rangle, 
\label{eq:occ_matrix_0}
\end{equation}
where $v$ and $\sigma$ are the band and spin labels of the KS states, respectively, $\mathbf{k}$ indicates points in the first Brillouin zone (BZ), $f^\sigma_{v,\mathbf{k}}$ are the occupations of the KS wavefunctions, and $\phi^I_{m}(\mathbf{r}) \equiv \phi_{m}(\mathbf{r} - \mathbf{R}_I)$ are localized orbitals centered on the $I^{\text{th}}$ atom at the position $\mathbf{R}_I$.
It is important to recall that the choice of the projector functions exerts a strong influence on the numerical values of calculated Hubbard parameters and affects the prediction of materials properties~\cite{Timrov2020a, KirchnerHall2021, Wang2016}.
Besides the localized atomic orbitals $\phi$ appearing in Eq.~\eqref{eq:occ_matrix_0}, there are other types of projector functions that may provide a more system-specific description of orbital occupations at some expense of computational and conceptual simplicity.
Particularly noteworthy in this context are L\"owdin-orthogonalized atomic orbitals~\cite{Timrov2020a, Mahajan2021} as well as Wannier functions~\cite{O’Regan2010, Tesch2022, Carta2024} (e.g., maximally localized ones~\cite{Marzari1997, Marzari1999, Schnell2002, Qiao2023, Qiao2023a}). 
A more detailed discussion of these projector functions including specific advantages and drawbacks can be found in Ref.~\cite{Timrov2020a}.

To carry out practical calculations using the energy functional of Eq.~\eqref{eq:Edftu}, the Hubbard parameters $U$ and $V$ must be determined for all of the selected target manifolds.
The DFPT approach employed in this work evaluates these parameters based on the heuristic finding that Hubbard corrections can (\textit{locally}) eliminate the spurious deviations of the total energy from PWL ~\cite{Cococcioni2005, Zhao2016}:
\begin{equation}
    \label{eq:local-curvature}
    U^{II} = \left . \frac{\partial^2 E_{\mathrm{DFT}}}{\partial [\mathrm{Tr}(\mathbf{n}^{II})]^2} \right |_q  \,, \,
    \hspace{3cm} 
    V^{IJ} = \left . \frac{\partial^2 E_{\mathrm{DFT}}}{\partial [\mathrm{Tr}(\mathbf{n}^{IJ})]^2} \right |_q ,
\end{equation}
where $|_q$ means that the expressions shall be evaluated at a fixed total charge $q$ of the system and $\mathrm{Tr}(\mathbf{n}^{IJ})$ is the trace of the occupation matrix $\mathbf{n}^{IJ}$ (where $\mathbf{n}^{IJ} = \sum_\sigma \mathbf{n}^{IJ\sigma}$, and $\mathbf{n}^{IJ\sigma}$ is the matrix whose elements are $n^{I J \sigma}_{m m'}$), whose elements are obtained from Eq.~\eqref{eq:occ_matrix_0}.
Because a direct control of orbital occupations is not always tractable, instead of computing the response of $E_{\mathrm{DFT}}$ to changes in the occupation matrix, one can instead compute the response of the occupation matrices to a linear perturbation $\alpha^J$~\cite{Cococcioni2005}, 
\begin{equation}
    \chi_0^{IJ} = \frac{\partial [\mathrm{Tr}(\mathbf{n}^{II})_{0}]}{\partial\alpha^{J}} \hspace{2cm}
    \chi^{IJ} =  \frac{\partial [\mathrm{Tr}(\mathbf{n}^{II})]}{\partial\alpha^{J}}~,
    \label{eq:chi}
\end{equation}
where $\chi_0$ and $\chi$ are the bare and self-consistent response matrices, respectively. 
The former quantity is computed before the self-consistent readjustments of the Hartree and xc potentials due to the perturbation, whereas the latter is obtained at self-consistency of the DFPT calculation~\cite{Timrov2018}.
We note that in order to derive Hubbard $V^{IJ}$ parameters consistent with Eq.~\eqref{eq:local-curvature}, in Eq.~\eqref{eq:chi} one should use the responses of $\mathrm{Tr}[\mathbf{n}^{IJ}]$ instead of $\mathrm{Tr}[\mathbf{n}^{II}]$.
However, the current \QE implementation of DFPT relies on Eq.~\eqref{eq:chi} and we expect this inconsistency to have a negligible influence on the numerical values of the resulting $V^{IJ}$ parameters.
With the response matrices from Eq.~\eqref{eq:chi}, the Hubbard parameters $U$ and $V$ can be computed according to Refs.~\citenum{Cococcioni2005, Jr2010}:
\begin{equation}
    U^I = \left(\chi_0^{-1} - \chi^{-1}\right)^{II} 
    ~,
    \quad 
    V^{IJ} = \left(\chi_0^{-1} - \chi^{-1}\right)^{IJ}
    ~.
    \label{eq:Ucalc}
\end{equation}
We note in passing that the ``full'' inversion of the $\chi$ and $\chi_0$ matrices as practiced in Eq.~\eqref{eq:Ucalc} and used throughout this work is not the only way of computing Hubbard parameters, and that other possibilities have been explored~\cite{Linscott2018, Moore2024}.
Particularly for manifolds that respond to perturbations very weakly (such as $d^{10}$ ions) or that display a strong intra-shell screening~\cite{Macke2024}, the linear-response approach presented here can result in oscillating or diverging results and must be used with great care~\cite{Kulik2008, Yu2014}.

Within the DFPT formalism, the linear responses of Eq.~\eqref{eq:Ucalc} can be conveniently expressed in terms of monochromatic perturbations modulated with wave vectors $\mathbf{q}$ as~\cite{Timrov2018}:
\begin{equation}
    \frac{\partial n^{I\sigma}_{mm'}}{\partial\alpha^{J}} 
    \equiv
    \frac{\partial n^{sl,\sigma}_{mm'}}{\partial\alpha^{s'l'}} 
    =
    \frac{1}{N_{\mathbf{q}}}
    \sum_{\mathbf{q}}^{N_{\mathbf{q}}}
    e^{i\mathbf{q}\cdot(\mathbf{R}_{l} - \mathbf{R}_{l'})}
    \Delta_{\mathbf{q}}^{s'} n^{s \sigma}_{mm'}
    ~,
    \label{eq:dnq}
\end{equation}
where $s$ and $s'$ indicate the atomic indices within the unit cell, while $l$ and $l'$ are the unit cell indices, such that $I\equiv(sl)$ and $J\equiv(s'l')$, $N_\mathbf{q}$ is the total number of $\mathbf{q}$ points, $\Delta_{\mathbf{q}}^{s'} n^{s \sigma}_{mm'}$ is the lattice-periodic response of occupation matrices to the $\mathbf{q}$-specific perturbation, $\mathbf{R}_l$ and $\mathbf{R}_{l'}$ are the Bravais lattice vectors.
Further details can be found in Refs.~\cite{Timrov2018, Timrov2021}.
This approach allows to avoid using computationally expensive supercells, making it the method of choice for large-scale applications.
%

\section*{RESULTS}

\subsection*{Computational workflows and data structure}
\noindent In this section, we describe the computational workflow that automates the self-consistent calculation of Hubbard $U$ and $V$ parameters using DFPT.
\begin{figure}[t]
    \centering
    \includegraphics[width=0.95\textwidth]{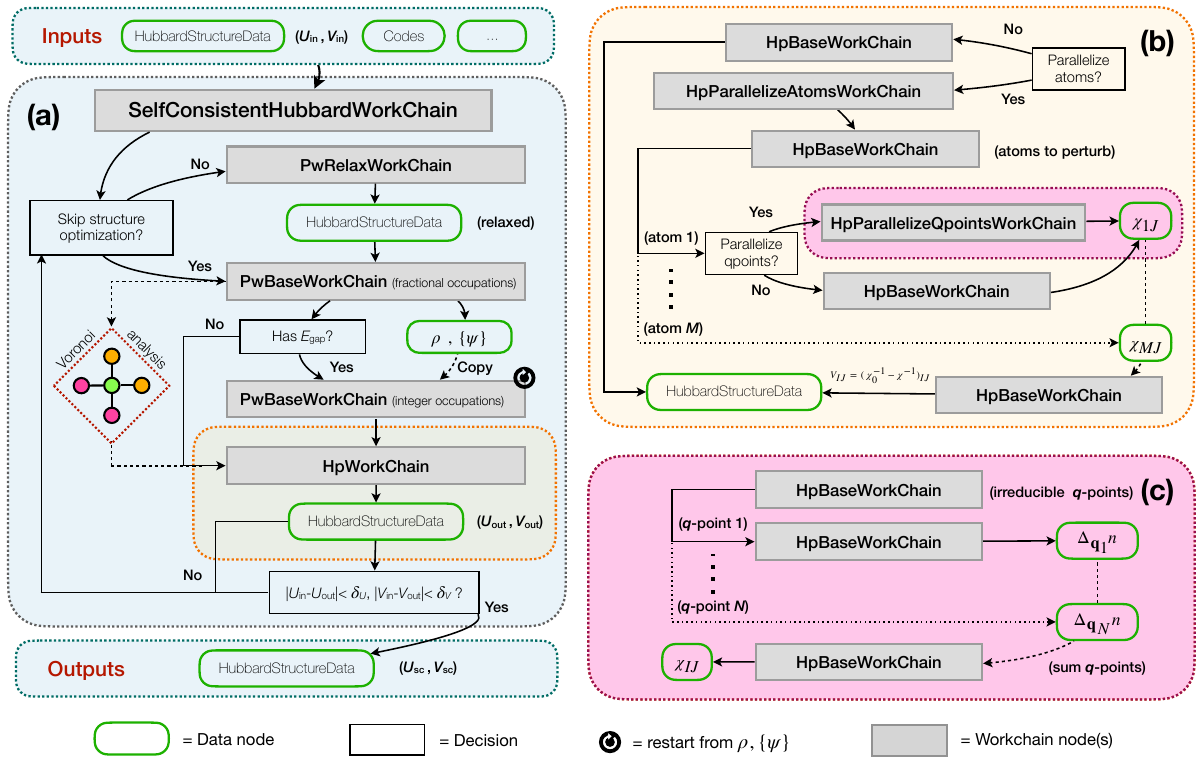}
    \caption{
        \textbf{Schematic illustration of the \texttt{aiida-hubbard} plugin that automates the self-consistent calculation of the Hubbard $U$ and $V$ parameters.}
        (a)~The main \texttt{Self\-Consistent\-Hubbard\-Work\-Chain} workflow which automates the self-consistent calculation of $U$ and $V$ parameters. It iterates (optionally) structural optimizations via the \texttt{PwRelaxWorkChain}, ground-state calculations via the \texttt{PwBaseWorkChain}, and the DFPT calculations of Hubbard parameters via the \texttt{HpWorkChain}. 
        In particular, the latter can be used to fully exploit the parallel capabilities of the \textsc{HP} code~\cite{Timrov2022}, i.e. by (optionally) first parallelizing the calculations over inequivalent Hubbard atoms to perturb, using the \texttt{HpParallelAtomsWorkChain} (panel (b)), and then (optionally) parallelizing over irreducible monochromatic perturbations ($\mathbf{q}$ points) via the \texttt{HpParallelQpointsWorkChain} (panel (c)). 
        These nested calls are visualized by the different colored boxes. 
    }
    \label{fig:workflow}
\end{figure}
%


\textbf{General structure of the aiida-hubbard plugin.} 
The workflow shown in Fig.~\ref{fig:workflow} contains several key building blocks.
%
The main self-consistent Hubbard workflow and its subprocesses are implemented as AiiDA \textit{workchains} \cite{Uhrin2021}, powering the automated handling and reproducibility of all the calculations. 
These workchains are represented by grey rectangles.
Data nodes, representing inputs and outputs of the workflows and calculations, are depicted by green rounded boxes.
For clarity, not the entire nested list of inputs is shown in the figure, but only the mandatory input data classes that are needed to run the workflow.
The light grey box (Fig.~\ref{fig:workflow}a) contains the outline of the \texttt{Self\-Consistent\-Hubbard\-Work\-Chain}, the main workflow of the package, which carries out the self-consistent calculation of the Hubbard parameters.
Its ``child'' processes are the \texttt{PwBaseWorkChain} and the \texttt{PwRelaxWorkChain}, which are specialized workchains that run the \textsc{PW} code (\texttt{pw.x} executable)~\cite{Giannozzi2009} of \QE as part of the \texttt{aiida-quantumespresso} plugin~\cite{Huber2021}, and the \texttt{HpWorkChain} managing the parallel capabilities of the \textsc{HP} code (\texttt{hp.x} executable)~\cite{Timrov2022} of \QE.
The orange and pink boxes (Fig.~\ref{fig:workflow}b and c) zoom in on the fine-grained parallelization facilitated by the DFPT framework and the \textsc{HP} code. 
The main input and output of the workflow is a \texttt{Hubbard\-Structure\-Data} object, a new data type created to store information on the Hubbard functional together with the atomistic structure.


\textbf{Joint description of atomistic structures and Hubbard interactions.} 
Hubbard corrections are defined within boundary conditions that extend beyond the mere numerical values of the interaction parameters (e.g., $U$ and $V$).
First and foremost, every set of Hubbard parameters used in a calculation is tied to the atomistic structure to which it is applied.
Therefore, \texttt{Hubbard\-Structure\-Data} unifies the description of Hubbard corrections and the respective atomistic structures in one class, and provides auxiliary user-friendly utilities that facilitate the initialization and handling of Hubbard-related data.
This is achieved by combining the structural information, inherited from the \texttt{StructureData} class already available in AiiDA, with a new \texttt{Hubbard} class presented below.

In \texttt{Hubbard}, we distinguish three key components: the mathematical formulation of the correction (flavor), the Hubbard projectors, and the interaction parameters. 
The Hubbard formulation (e.g., ``Dudarev''~\cite{Dudarev1998} or ``Liechtenstein''~\cite{Liechtenstein1995}) and the kind of projectors (e.g., ``atomic", ``ortho-atomic"~\cite{Timrov2020a}) are specified as strings, whereas the interaction parameters are stored as a list of instances of \texttt{HubbardParameters}.
We decide not to include the pseudopotentials in \texttt{Hubbard} since, thanks to the provenance model of AiiDA, this piece of information can be easily traced back or simply added as extra metadata, thus avoiding inefficient repetition of data. Nevertheless, we recall the impact of pseudopotential choice (i.e., of the choice of projectors) on the numerical value of Hubbard parameters computed using LR-cDFT, which can vary by $2-3$~eV~\cite{Kulik2008}.

\begin{figure}[t]
    \centering
    \includegraphics[width=0.475\textwidth]{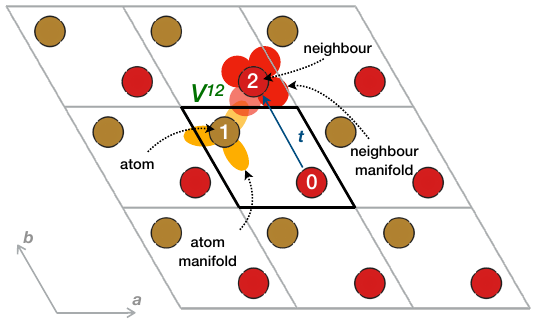}
    \caption{
        \textbf{Two-dimensional periodic system summarizing the quantities needed to describe a Hubbard interaction $V^{IJ}$.}
        This example shows a 2D lattice with two crystallographically inequivalent atoms in the unit cell, indexed 0 and 1. 
        To store a Hubbard interaction between atoms 1 and 2, where 2 is a periodic image of atom 0, one stores the indices 1 and 0 along with the translation vector $\mathbf{t}$ that maps atom 0 onto atom 2 using integer multiples of the cell vectors $\mathbf{a}$ and $\mathbf{b}$ (in this case, $\mathbf{t} = (0,1)$).
        Moreover, the yellow and blue orbitals shown around atoms 1 and 2 indicate the target Hubbard manifolds, which must also be stored.
        Finally, $V^{12}$ represents the value of the interaction parameter in energy units, usually expressed in eV.
    }
    \label{fig:hubbard_interaction}
\end{figure}

\texttt{HubbardParameters} is an extra class defining a single Hubbard interaction that contains its type ($U$, $V$, $J$ etc.), the indices and manifolds (e.g. $3d$, $2p$, etc.) of the atom(s) involved, as well as the value of the respective parameter expressed in energy units (eV).
To allow for the description of interactions between two distinct atoms (intersites), \texttt{HubbardParameters} additionally stores a second atomic index and a second target manifold.
For calculations with periodic boundary conditions, intersite couples might be located in periodic images of the unit cell. 
Therefore, the first atomic index is always referenced within the unit cell, while every second atomic index is augmented by a translation vector $\mathbf{t}$ pointing to the atom's corresponding periodic image.
For instance, for the structure shown in Fig.~\ref{fig:hubbard_interaction}, an intersite interaction between atoms 1 and 2 ($V^{12}$) would be stored as an interaction between atoms 1 and 0 plus the translation vector that maps atom 0 onto atom 2. 
The $U$ and any other onsite interaction parameters can be defined by specifying the same index and manifold for both fields, and by assigning a null translation vector $\mathbf{t} = \mathbf{0}$.
%

\textbf{Choosing the interacting Hubbard couples.} 
Before intersite Hubbard $V^{IJ}$ parameters can be evaluated and applied, it is necessary to define the interaction couples by providing their atomic indices and target manifolds. 
Since intersite corrections are intended for systems where orbital hybridization plays an important role~\cite{Jr2010}, $V^{IJ}$ parameters are typically established for nearest-neighbor couples  (e.g., between the \textit{d}-shell of a central TM atom and the \textit{p}-shells of its ligands).
In practice, it is desirable to let the user specify the elements to be considered as interacting, while delegating the (potentially error-prone) determination of the respective atomic indices $I$ and $J$ to an algorithm.
However, generally the search for nearest neighbors can be challenging, particularly in structures hosting simultaneously different coordination environments (e.g., tetrahedral and octahedral  sites in spinels).
Counting the neighbors contained in a sphere around each central atoms offers a straightforward solution but necessitates a common radial cutoff value that must be at the same time large enough to include all of the specified couples but small enough not to introduce additional interactions. 
Such a common cutoff radius can be hard to determine, or might not even exist at all; for instance in amorphous and low-symmetry ordered structures, where coordination environments are notoriously difficult to characterize. 
An additional problem occurs in workflows involving structural optimizations: when the cutoff is recalculated following a structural relaxation, it cannot be guaranteed that the same (and only the same) atoms are contained in the new sphere. 
These issues become particularly unmanageable in high-throughput calculations, thus motivating the need for a robust automation of the process, which should be carried out in each iteration of the self-consistent cycle.
In \texttt{aiida-hubbard}, the analysis of nearest-neighbours is therefore performed using the Voronoi tessellation method~\cite{Blatov*2004} as implemented in the \texttt{Pymatgen} core utilities~\cite{Ong2013, Isayev2017}.
This parameter-free approach systematically accounts for diverse coordination environments, even if these coexist in the same atomistic structure, without the need for a common radial cutoff.


\textbf{Description of the Self\-Consistent\-Hubbard\-Work\-Chain.}
Having established a consistent data structure for storing Hubbard data and having automated the determination of intersite Hubbard couples, we now present the core workflow of \texttt{aiida-hubbard}.
The \texttt{Self\-Consistent\-Hubbard\-Work\-Chain} combines the capabilities of the \texttt{Pw\-Base\-Work\-Chain} and \texttt{Pw\-Relax\-Work\-Chain}~\cite{Huber2020,Huber2021} with the \texttt{HpWork\-Chain}.
The self-consistency of the Hubbard parameters~\cite{Cococcioni2019, Timrov2021} is achieved iteratively by performing (i) structural optimizations, (ii) single-point DFT+$U$+$V$, and (iii) DFPT calculations of $U$ and $V$ until convergence. 
After each structural optimization, the relaxed structure is used to perform a single-point DFT+$U$+$V$ calculation with fractional electronic occupations (indicated by ``smearing" in Fig. \ref{fig:workflow}a) in order to identify whether the system is metallic or insulating.
If the electronic structure displays a finite band gap, an extra calculation with fixed integer occupations is performed (indicated by ``fixed" in Fig. \ref{fig:workflow}a), which reuses the previously obtained charge density and wavefunctions in order to accelerate convergence and to preserve the determined magnetic ground state in case of spin-polarized calculations.
This second single-point step is fundamental to avoid numerical divergence in the DFPT calculation at $\mathbf{q}=\mathbf{0}$~\cite{Baroni2001, Timrov2022}.
Finally, the DFT+$U$+$V$ ground-state is used to carry out the DFPT calculation that predicts the new set of Hubbard parameters.
After completing a cycle, these Hubbard parameters are then used for the next iteration.
This sequential procedure is repeated until the variations in parameters fall below user-predefined thresholds $\delta U$ and $\delta V$ (typically in the range of 0.01 to 0.1~eV).

We note that other ways of conducting the self-consistency procedure are also possible.
For instance, the structural optimization can be omitted so that the Hubbard parameters are converged or at least iterated a couple of times for a fixed atomistic structure. 
Alternatively, an intermediate strategy could be pursued in which structural optimizations are not performed at every cycle, but instead intermittently (e.g., only once every 3-5 iterations).
Another potentially useful approach that might reduce the number of iterations involves using a reasonable guess for the Hubbard parameters instead of starting from the initial values $U_\mathrm{in} = V_\mathrm{in} = 0$.
Initial values can either stem from a machine-learning model~\cite{Uhrin2025} or can be chosen empirically.
For very oscillating Hubbard parameters, a mixing strategy can be introduced. The origin of the oscillations can be attributed in part to the $dU/d\mathbf{R}$ contribution (usually discarded) to the forces, which can have a sizeable effect on structural relaxation~\cite{Kulik2011b}.

%


\textbf{Parallelization levels of the HpWorkChain.}
A crucial aspect for practical applications of the \texttt{Self\-Consistent\-Hubbard\-Work\-Chain} is the computational time required to complete an iteration.
As the most demanding step of each cycle generally consists in the computation of the Hubbard parameters with DFPT, finding strategies to accelerate the latter is desirable.
While the \textsc{HP} code of \QE provides several options that allow for a distribution of the computational load, some of these parallelization levels can be coordinated by a high-level orchestrator.
For this purpose, \texttt{aiida-hubbard} implements the \texttt{HpWorkChain}, which allows the user to parallelize the DFPT calculations in an automated fashion using up to two levels of parallelization.
The first layer, a parallelization over atoms, arises because each element $IJ$ of the response matrix $\chi$  can be computed independently from the others (see Eq. \eqref{eq:Ucalc}).
This functionality is provided by the \texttt{Hp\-Parallelize\-Atoms\-Work\-Chain} (see Fig.~\ref{fig:workflow}b).
The second level of parallelization can be achieved using the \texttt{Hp\-Parallelize\-Qpoints\-Work\-Chain} (Fig.~\ref{fig:workflow}c), which distributes the calculation of the independent wavevectors $\textbf{q}$ that contribute to the total occupation matrix (see Eq.~\eqref{eq:dnq}).
Particularly for large systems, leveraging both strategies can offer a computational boost on massively parallel architectures, where the DFPT calculations for each perturbed Hubbard atom and \textbf{q}-point can be executed concurrently on the available compute nodes.
The independent DFPT calculations spawned by the parallel workchains are managed by the \texttt{Hp\-Base\-Work\-Chain}, a ``base'' workflow designed to run the \texttt{hp.x} binary of the \textsc{HP} code featuring automated submission, retrieval and error handling capabilities.
Errors are addressed effectively by modifying the inputs without any user intervention, which is crucial for high-throughput calculations.
For example, if the self-consistent response does not converge within the maximum number of iterations, \texttt{Hp\-Base\-Work\-Chain} submits a new \textit{hp.x} job with a lower mixing factor the response charge density mixing needed for solving iteratively the Sternheimer equations of DFPT~\cite{Timrov2018, Timrov2021}.
%


\textbf{Semi-automatic input preparation.} 
In the preceding sections, we have conceptualized workflows to perform self-consistent calculations of Hubbard parameters. 
However, the results of both the workflows and the individual DFT+$U$+$V$ and DFPT calculations depend upon a large number of inputs. 
These inputs comprise code-specific keywords such as convergence thresholds (on energy, forces, stresses), cutoff values, mixing parameters, $\mathbf{k}$-point grids for the Brillouin zone sampling, pseudopotentials, but also metadata associated with the computational resources including the walltime limit and the number of computational nodes and cores requested, to name a few.
Not only for non-expert users, choosing suitable values for all of the inputs and generating (syntactically correct) input files can be a tedious and error-prone task.
To reduce this complexity, \texttt{aiida-hubbard} features a \texttt{get\_builder\_from\_protocol} method for each of the workchains~\cite{Huber2021,Bastonero2024}.
This method automatically populates the inputs, while the user is left with the task of providing only three remaining indispensable pieces of information: (i) an instance of \texttt{Hubbard\-Structure\-Data} (i.e. the atomistic structure with initialized Hubbard parameters), (ii) AiiDA code instances containing information on how to run the \textsc{PW} and \textsc{HP} codes \cite{Huber2020,Uhrin2021}, and (iii) a string defining in a general fashion the accuracy of the calculation called \texttt{protocol} (``fast'', ``moderate'', or ``precise''). 
A summary of the main calculation parameters these protocols \textit{initialize} is reported in Supplementary Table 1.
Importantly, after calling the \texttt{get\_builder\_from\_protocol} method, the user receives a pre-populated set of inputs, which can then be checked and modified before being used for the execution of the workflow.

\subsection*{Impact of structural optimizations on self-consistent Hubbard parameters}
In the outline of the \texttt{Self\-Consistent\-Hubbard\-Work\-Chain}, we have presented different strategies for the self-consistent computation of Hubbard parameters.
Hence, it is worthwhile to investigate the impact of different schemes on the numerical value of the Hubbard parameters.
Here, we perform numerical experiments in which we compare two commonly employed approaches: $(i)$ converging the Hubbard parameters by alternating single-point DFT+$U$+$V$ and DFPT calculations, and $(ii)$ by performing the optimization of the lattice parameters and the atomic positions at the beginning of each cycle.
In the following we will refer to strategy (i) as \textit{NR scheme} (no relaxation) and (ii) as \textit{FR scheme} (full relaxation).
For the sake of simplicity, we conduct these experiments by computing only the self-consistent Hubbard $U$ parameters, neglecting intersite $V$ interactions.
We apply these two schemes to six chemically and structurally diverse crystalline solids containing Li and Fe that have been investigated experimentally~\cite{Huber2022}.
We apply onsite $U$ corrections to the Fe-$3d$ states, and converge the $U$ parameter within $\delta U = 0.1\,$eV.

\begin{figure}[t]
    \centering
    \includegraphics[width=0.95\textwidth]{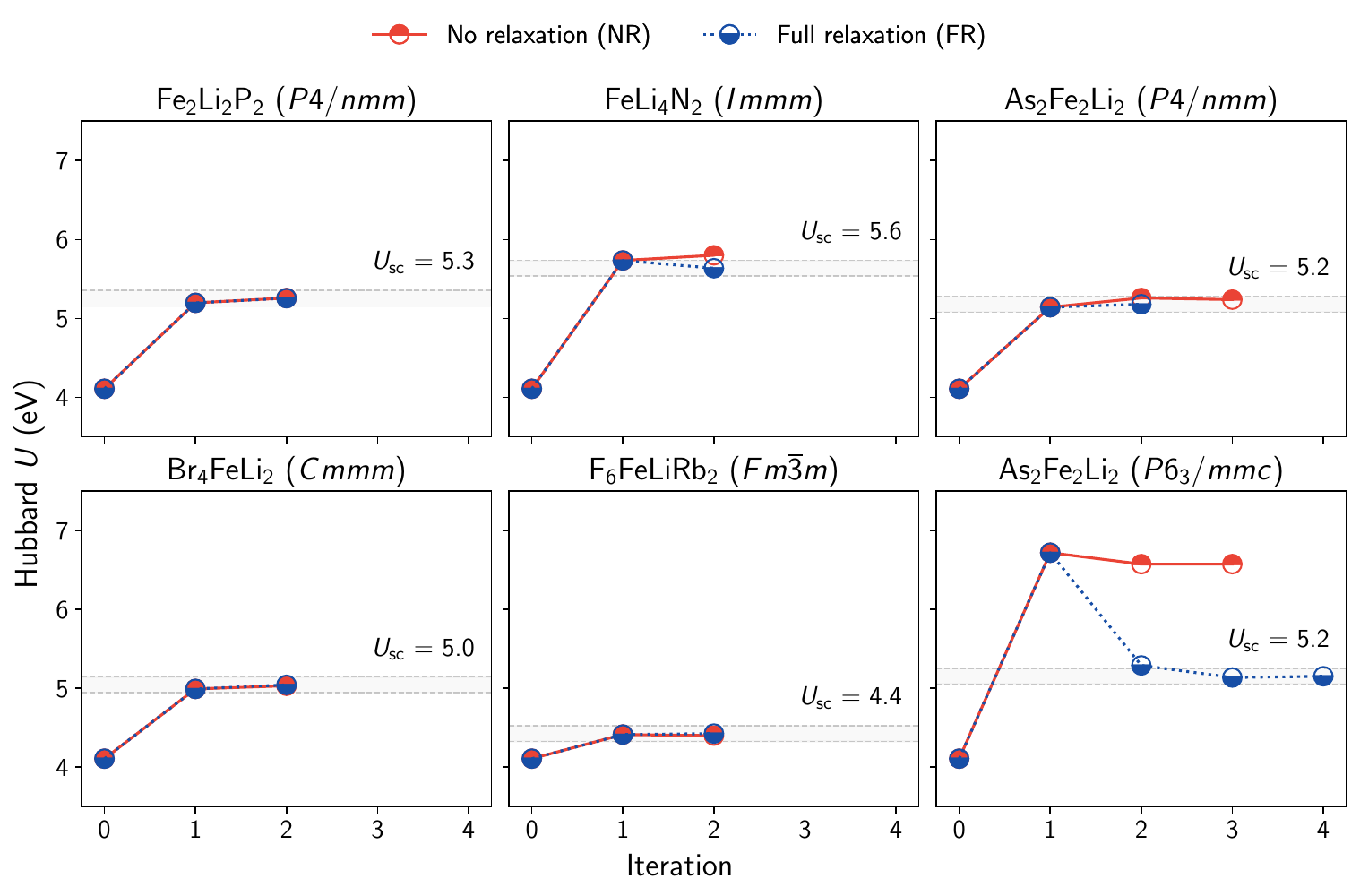}
    \caption{
        \textbf{Self-consistent convergence of Hubbard $U$ for Fe-$3d$ using two different schemes.}
        Values of Hubbard $U$ for six different bulk Fe-containing solids as a function of the iteration of the self-consistent cycle,  using the NR (red) and FR (blue) schemes.
        For each compound, the associated Hill formula and space group is reported in the title of the corresponding panel.
        Iteration 0 corresponds to the starting guess $U_{\text{Fe}} = 4$~eV. Geometry optimizations of the FR cycle were omitted in iteration 1.
        The gray shaded area shows the convergence range $U_{\text{sc}} \pm \delta U$ ($\delta U = 0.1$~eV), around the final values of the FR cycles. 
    }
    \label{fig:self-consistencies}
\end{figure}

Figure~\ref{fig:self-consistencies} shows the computed numerical values of $U$ during the self-consistency cycle, with red and blue half-filled dots referring to the NR and FR schemes, respectively.
The data points at iteration 0 represent the starting guess $U_{\text{Fe}}=4$~eV that was used to initialize the self-consistent cycles.
We find that the numerical values of $U$ computed at iteration 1 are always identical for both approaches since the geometry optimizations of the FR scheme were omitted at this iteration.
It can be observed that the final self-consistent values of $U$, reported as $U_{\text{sc}}$, depend on the specific compound and range from 4.4 to 5.6~eV.
Interestingly, none of the two approaches consistently outperforms the other with respect to the number of iterations required to converge $U$. 
Furthermore, the two strategies yield the same $U_{\text{sc}}$ values for all but one structure (\ce{As2Fe2Li2} with the space group $P6_3/mmc$), where the $U_{\text{sc}}$ parameter obtained for the NR approach exceeds that of the FR approach by about 1.4~eV. 
This observation can be explained with a peculiar volume expansion by over 150\% upon optimization of the crystal structure, which concurs with significant changes in the electronic structure.  
In fact, the total projected occupation of the Fe-$3d$ shell decreases by about 1$e^-$ following the volume expansion.
Since the Hubbard parameters are calculated from the response of the occupation matrices, the change of the latter leads to a shift in the computed $U$ value.
The large increase in volume results from the initial experimental structure being measured under high-pressure conditions ($P > 1000\,$kPa)~\cite{Huber2022}.
All of the other structures display less pronounced volume deviations (14\% maximum and 6\% on average), and also present only negligible changes in the occupations of the Fe-$3d$ manifold. 
Thus, in these cases, the FR and NR strategies yield the same $U_{\text{sc}}$ parameters (within the chosen threshold).
A similar observation can be made in an analogous numerical experiment carried out with Mn-bearing compounds, whose results are presented in Supplementary Figure 1.

Therefore, the impact of structural optimization on self-consistent Hubbard parameters is negligible for most of the materials considered here.
However, this should not be assumed as a general trend, as the importance of the geometry optimization depends on the magnitude of rearrangements between the final and starting atomistic structures.

\subsection*{Trends in Hubbard parameters from 105 Li-bearing materials}
Having demonstrated the flexibility of the workflows to account for different self-consistent strategies, we now proceed to showcase their scalability and robustness by carrying out calculations of Hubbard $U_{\text{sc}}$ and $V_{\text{sc}}$ parameters across a diverse set of materials.
While several mid- and high-throughput studies on the prediction of Hubbard parameters can be found in literature~\cite{KirchnerHall2021, Xiong2021, Tesch2022, Gelin2024, Moore2024}, these were limited to the prediction of $U$ (and sometimes $J$) parameters only, not to mention computational difficulties encountered due to the supercell approach required by LR-cDFT in some of these studies. 
The present study focuses on 115 experimentally known crystalline solids containing Li, Fe or Mn, and not limiting additional elements, with unit cells of 32 atoms or fewer.
The full list of materials studied is presented in Supplementary Table 2.
These compounds are relevant as they represent potential candidates for cathodes in novel Li-ion batteries. 
Moreover, it has been shown that the DFT+$U$+$V$ approach can accurately predict various properties of such materials, including open-circuit voltages~\cite{Cococcioni2019, Timrov2022a, Timrov2023}.

We compute the onsite $U$ parameters for the TM $3d$ shells, and consider intersite $V$ interactions between the TM $3d$ shells and the $p$ shells of neighboring chalcogenide atoms (O, S, Se, Te).
This choice is based on the expectation that inter-atomic orbital hybridization (e.g., the formation of $\sigma_{p-d}$ states) is most pronounced for these couples.
Due to the absence of chalcogenides in 42 of the 115 structures, only the onsite $U$ parameter is computed for these cases.
To not only obtain the self-consistent Hubbard parameters but also the structural DFT$+U(+V)$ ground states, we leverage the FR scheme presented in the previous section, initializing the workflows with $U_{\text{TM}} = 5.0$~eV and $V=0$~eV, when applicable.
All but ten of the submitted \texttt{Self\-Consistent\-Hubbard\-Work\-Chain} processes finished successfully, managing the automated recovery of several computational errors that occurred in about 6\% of the DFT+$U$+$V$ and DFPT calculations submitted.
The self-consistent cycles converged within 2.9 iterations on average. 
Thus, about three structural optimizations, single-point DFT$+U\!+\!V$ calculations, and DFPT runs are needed to converge the Hubbard parameters within the $\delta U = 0.1$ and $\delta V = 0.1$~eV thresholds.
In more detail, 34 workchains converged in 2 iterations, 58 workchains needed 3 iterations, and only 13 required 4 or more iterations to reach self-consistency.
Among the unsuccessful calculations using the workflow, many failed due to crashes of the \textsc{PW} and/or \textsc{HP} simulations caused by non-trivial numerical issues (e.g., with the minimization algorithms).
While future updates to the \QE distribution or adjustments in \texttt{aiida-quantum\-espresso} and \texttt{aiida-hubbard} may address these issues, we identified two compounds where convergence of the Hubbard parameters could not be achieved for physical reasons. 
We examine these cases in more detail in the Supplementary Discussion and Supplementary Table 3.

\begin{figure}[t]
    \centering
    \includegraphics[width=0.95\textwidth]{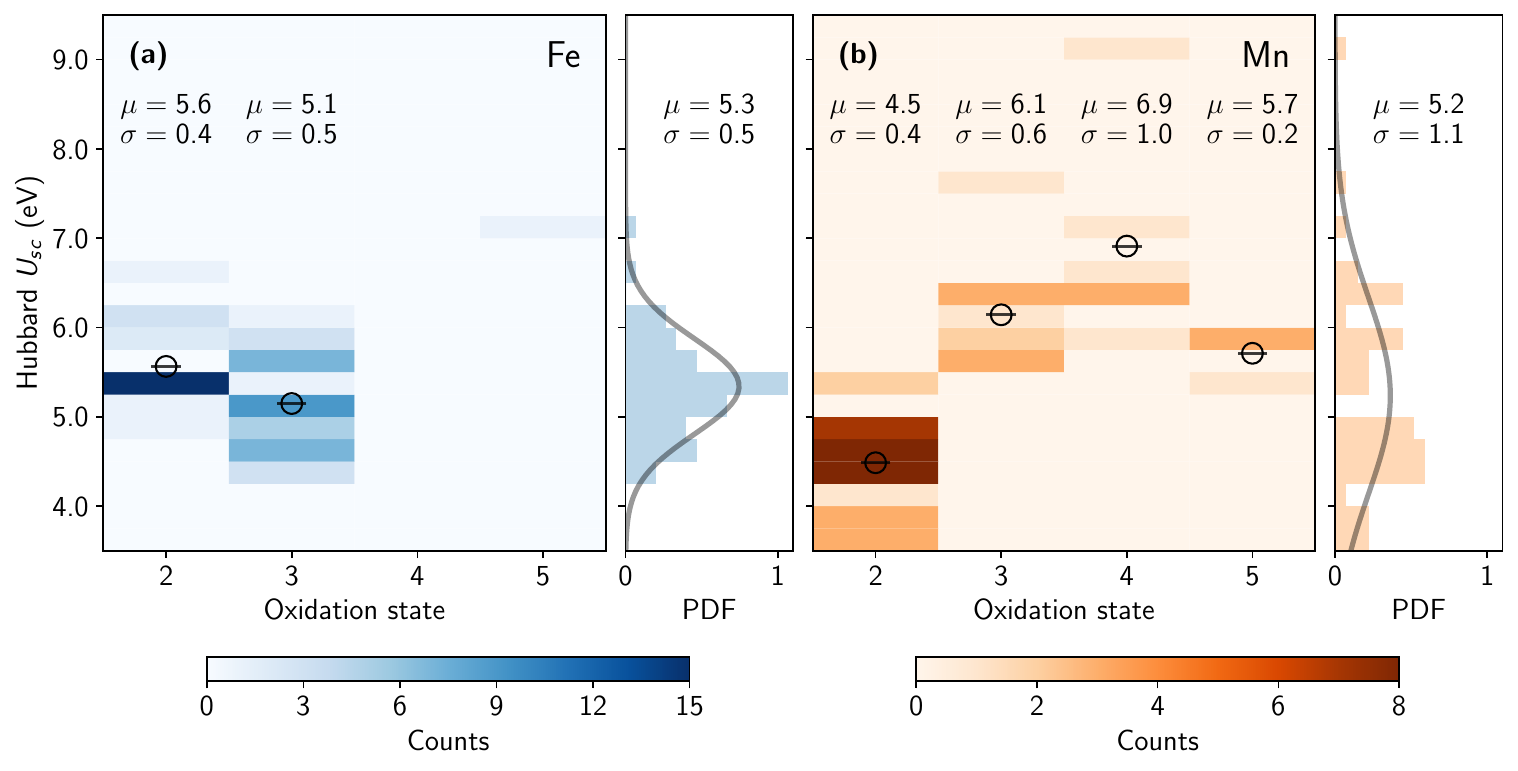}
    \caption{
        \textbf{Distributions of self-consistent Hubbard $U_{\text{sc}}$ parameters among 105 Li-bearing materials.}
        \textbf{(a,b)} Values of onsite Hubbard $U_{\text{sc}}$ parameters as a function of the OS of Fe and Mn.
        The side panels show the data probability distribution along with the fitted Gaussian distributions (gray lines) of $U_{\text{sc}}$ across the explored oxidation states and report the mean values ($\mu$) and standard deviations ($\sigma$) in units of eV.
    }
    \label{fig:hubbardsU}
\end{figure}

Figure~\ref{fig:hubbardsU} shows the range of self-consistent Hubbard parameters determined by the workchains, showing the dependence of $U_{\mathrm{sc}}$ on the OS of the TM elements.
The OSs are determined following the approach of Ref.~\cite{Sit2011} using a threshold of 0.8, i.e., only those eigenstates of the occupation matrix $\mathbf{n}^{II\sigma}$ whose eigenvalues $\lambda^\sigma$ are determined to be larger or equal to 0.8 are counted as occupied orbitals that determine the OS.
%
%
Since \ce{Fe} and \ce{Mn} are multivalent elements, the OSs they exhibit vary depending on the specific compound.
In the vast majority of cases Fe is present as \ce{Fe^{2+}} or \ce{Fe^{3+}}, whereas the rarer \ce{Fe^{5+}} specie was identified only for one material in our dataset, namely FeLa$_2$LiO$_6$ (c.f. ICSD entry \texttt{252554} and MaterialsCloud ID \texttt{mc3d-47750/pbe-v1}). 
Conversely, \ce{Mn} displays more variety.
Compounds were found for all OSs between $+2$ and $+5$ ($+7$ in one material, not shown in the figure; see Supplementary Table 3), with the majority of cases corresponding to \ce{Mn^{2+}}.
Interestingly, the mean values of the onsite $U_{\mathrm{sc}}$ parameters for \ce{Fe} and \ce{Mn} are close between each other, respectively $5.3$ and $5.2\,$eV, and are within the typical range of empirical Hubbard $U$ parameters used in the literature~\cite{Geatches2012, Hegner2016, Kim2022}.
However, Fig.~\ref{fig:hubbardsU} clearly illustrates that a universally transferable Hubbard parameter does not exist, not even when identical Hubbard projector functions are used, as in this study.
In fact, it can be seen that $U$ significantly depends on the OS of the TM ion. 
For instance, the average value of $U_{\mathrm{sc}}$ for \ce{Mn^{2+}} is 4.5\,eV, whereas that of \ce{Mn^{4+}} amounts to 6.9\,eV.
This observation is further supported by the fact that the data distributions do not show a symmetric Gaussian distribution 
of the $U$ values, especially when compared to the fitted Gaussian probability distribution functions (reported as gray lines), and instead indicate a clustering due to the distinct OSs, especially for \ce{Mn}.
Nevertheless, the wide range of $U$ values found even for compounds with identical OSs suggests that there must be other factors playing an important role. 
\begin{figure}[t]
    \centering
    \includegraphics[width=0.95\textwidth]{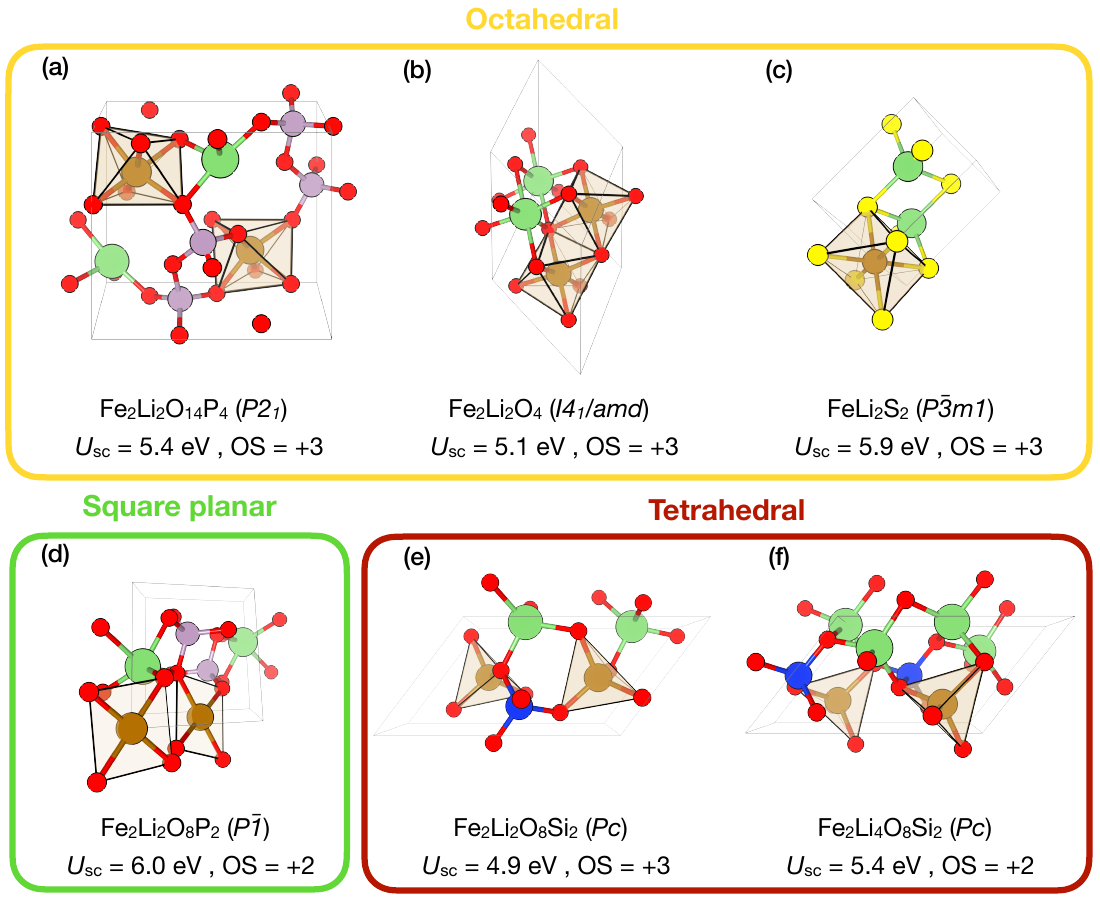}
    \caption{
        \textbf{A selection of structurally and chemically diverse compounds containing both Fe and Li.}
        (a-f) Images of the unit cells along with the corresponding Hill formulae, space group symbols in parenthesis, the final self-consistent Hubbard parameters $U_{\text{sc}}$, and the OSs of the \ce{Fe} ions.
        The images were rendered using VESTA~\cite{Momma2008}. 
        Color code: Fe (brown), O (red), Li (green), Si (blue), P (purple). 
    }
    \label{fig:structures}
\end{figure}

We therefore inspect a few representative structures in more detail and present them in Fig.~\ref{fig:structures}.
In particular, we focus on six ferric or ferrous compounds that vary not only with respect to the OS ($+2$ or $+3$) but also display chemically and structurally distinct coordination environments around the \ce{Fe} central atom.
Analyzing Fig.~\ref{fig:structures}, one can distinguish the following cases:
\begin{itemize}
  \item \textit{Different coordination environment, same OS, same ligand species}: The \ce{Fe} atoms in couples (a) and (e) as well as (d) and (f) display the same OSs and their ligand fields are composed of oxygen atoms, however, their coordination geometries differ. This alone leads to significant variations in $U$. For instance, $U=5.4\,$eV for the tetrahedrally coordinated \ce{Fe} in \ce{Fe2Li4O8Si2}, whereas $U$ amounts to $6.0\,$eV in \ce{Fe2Li2O8P2}, where \ce{Fe} exhibits a square planar coordination geometry. Hence, here we observe the variation in $U$ of 0.5--0.6~eV.
  \item \textit{Same coordination environment, same OS, different ligand species}: In all of (a), (b) and (c), the \ce{Fe} atoms are octahedrally coordinated \ce{Fe^{3+}} species. However, while in (a) and (b) the ligands are \ce{O} atoms, (c) features \ce{S} ligands. This again leads to a pronounced increase in $U$ by 0.5--0.8~eV when comparing oxides to sulfides.
  \item \textit{Same coordination environment, different OS, same ligand species}: Compounds (e) and (f) are structurally identical except that (f) contains two additional \ce{Li} atoms which reduce the OS of (f) from $+3$ to $+2$. A difference in $U$ of $0.5\,$eV is consistent with the previously discussed data shown in Fig. \ref{fig:hubbardsU} and earlier first-principles studies~\cite{Cococcioni2019, Timrov2022a, Timrov2023}.
  \item \textit{Same coordination environment, same OS, same ligand species}: Finally, in spite of the fact that (a) and (b) both posses octahedrally coordinated \ce{Fe^{3+}} ions, their $U$ values still differ by $0.3$~eV. Generally, these relatively moderate variations can be attributed to local distortions from the perfect $O_h$ point symmetry, which can be induced by Jahn-Teller effects and other kinds of distortion modes. However, since \ce{Fe^{3+}} is not Jahn-Teller active, these variations in $U$ can be attributed to variations in the volume of the FeO$_6$ octahedra induced by the surrounding atoms in the crystal structure and overall changes in the electronic screening of the Fe-$3d$ states by those atoms.
\end{itemize}
This simple and intuitive analysis highlights the strong dependence of $U$ on the local environment (in both structural and chemical terms) as well as on the OS of the central atom. 
In the case of Fe-bearing compounds, these effects lead to variation in $U$ less than 1~eV. 
However, in other compounds like those containing Mn, these variations can have a larger magnitude (see Fig.~\ref{fig:hubbardsU}).
\begin{figure}[t]
    \centering
    \includegraphics[width=0.95\textwidth]{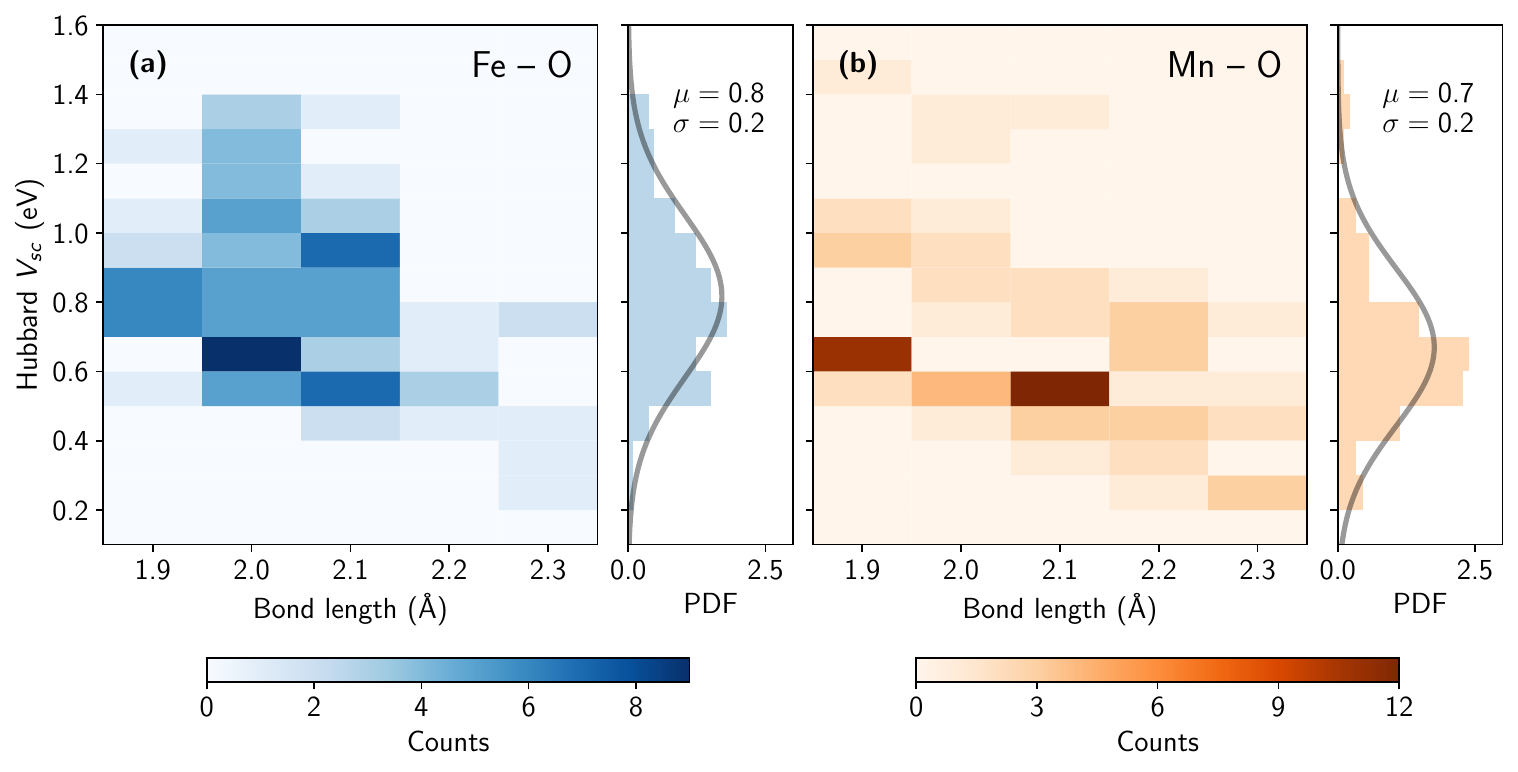}
    \caption{
        \textbf{Distributions of self-consistent Hubbard $V_{\text{sc}}$ parameters among 105 Li-bearing materials.}
        \textbf{(a,b)} Dependence of intersite Hubbard $V_{\text{sc}}$ parameters computed for the \ce{Fe $3d$\bond{-}O $2p$} and \ce{Mn $3d$\bond{-}O $2p$} interactions on the bond length between atoms.
        The side panels show the probability distribution functions (PDF) along with the fitted Gaussian distributions (gray lines) of $V_{\text{sc}}$ across the explored bond lengths and report the mean values ($\mu$) and standard deviations ($\sigma$) in units of eV.
    }
    \label{fig:hubbardsV}
\end{figure}
%

Given the clear connection between the onsite $U$ parameters and the local environments of the Hubbard atoms, it is intriguing to explore whether the intersite $V_{\mathrm{sc}}$ parameters display similar dependencies.
Since $V_{\mathrm{sc}}$ parameters are computed for atom pairs, their values might intuitively vary with the inter-atomic distance (bond length) between the interacting partners. 
Figure~\ref{fig:hubbardsV} shows the distribution of Hubbard $V_{\mathrm{sc}}$ parameters as a function of inter-atomic distance for \ce{Fe-O} (panel (a)) and \ce{Mn-O} (panel (b)) pairs, with corresponding PDFs displayed in the side panels.
Other couples were excluded from this analysis as the majority of compounds in the database are oxides (i.e., 60 out of 70 structures containing chalcogenide atoms).
However, a list of all compounds and their corresponding Hubbard parameters, including $V_{\mathrm{sc}}$ for couples with chalcogenides other than \ce{O}, can be found in Supplementary Table 2.
Analyzing Fig.~\ref{fig:hubbardsV}, it can be seen that the distributions of $V_{\mathrm{sc}}$ parameters among the \ce{Fe-O} (a) and \ce{Mn-O} (b) couples are quite similar.
In general, $V_{\mathrm{sc}}$ decreases with increasing bond lengths and vice versa.
Notably, the $V_{\mathrm{sc}}$ parameters in Fig.~\ref{fig:hubbardsV} can be described by Gaussian distributions with mean values of $0.8\,$eV (\ce{Fe-O}) and $0.7\,$eV (\ce{Mn-O}), and a standard deviation of $0.2\,$eV for both couples. 
Nonetheless, there are significant variations in $V_{\mathrm{sc}}$ that cannot be explained by a bond length dependence alone. 
For example, $V_{\mathrm{sc}}$ values computed for \ce{Fe-O} interactions at bond lengths of about $2.0\,$\AA\: vary between 0.5 and $1.4\,$eV.
Moreover, in Fig.~\ref{fig:hubbardsV}b one might distinguish at least two clusters: one that starts at $V\approx 0.3\,$eV for bond lengths of $2.3\,$\AA\: and ends at $V\approx 0.6$~eV for distances of $2.3\,$\AA, and a second one ranging from $V\approx 0.6$~eV (at $2.3\,$\AA) to $V\approx 1\,$eV (at $1.9\,$\AA).
As for the onsite $U$, such substantial variations reflect the non-trivial dependence of $V$ parameters on both electronic and geometric degrees of freedom.
%

\section*{DISCUSSION}
We have presented \texttt{aiida-hubbard}, a computational framework based on the AiiDA infrastructure~\cite{Pizzi2016} that automates, in a reproducible yet flexible fashion, the self-consistent calculation of onsite Hubbard $U$ and intersite Hubbard $V$ parameters leveraging the parallel capabilities of DFPT\cite{Timrov2022}.
We devised \texttt{HubbardStructureData}, a FAIR~\cite{Wilkinson2016} code-agnostic data type that jointly encodes all information relevant to the Hubbard corrections and the atomistic structure.
The difficulty of automatically defining intersite $V$ interactions for a generic coordination environment has been solved by exploiting a Voronoi tessellation algorithm at runtime.
To showcase the workflow's capabilities, we employed it to compute Hubbard parameters for 115 Li-bearing bulk solids with potential relevance for electrochemical applications, including Li-ion batteries.
As evidenced by the success rate of 91\% (105 in 115 workchains), \texttt{aiida-hubbard} is robust and highly reliable due to its integrated and automatic error handling, especially when considering that the selected compounds were quite realistic featuring unit cells of up to 32 atoms, diverse chemistry and bonding environments. 
An analysis of trends in the distribution of self-consistent Hubbard $U$ parameters revealed a strong dependence of the $U$ values of $3d$ manifolds on the oxidation states of the respective TM atoms.
At the same time, the coordination environment (i.e., number, arrangement and kind of ligands) also exerts substantial influence on $U$. 
The Hubbard $U$ parameter of \ce{Fe} $3d$ manifolds varies by up to $\sim 3\,$eV, whereas for the more polyvalent \ce{Mn} larger variations in $U$ by $\sim 6\,$eV are observed.
In particular, we found characteristics shifts of the numerical value of $U$ upon change in oxidation state or coordination environment, respectively about 0.5~eV and 1.0~eV for Fe and Mn.
Variations in the intersite $V$ parameters were smaller, ranging between 0.2~eV and 1.6~eV when considering TM--oxygen pairs.
These values correlate with the bond lengths between the interacting atoms, generally decreasing in magnitude as the interatomic distance increases; 
however, significant fluctuations remain that could not be explained based on distance arguments alone.
These observations indicate that the numerical values of $U$ and $V$ are subject to a complex interplay between electronic and structural degrees of freedom, meaning that the parameters cannot be accurately determined based on oxidation state or coordination environment only.
Therefore, machine learning models designed for predicting Hubbard parameters will likely need to incorporate descriptors of both kinds (e.g., using the OS~\cite{Sit2011} and ACE~\cite{Drautz2019} methods) to enhance the predictive accuracy.
For this purpose, it would also be desirable to investigate and quantify the impact of the input parameters ($\mathbf{k}$- and $\mathbf{q}$-point meshes, energy cutoffs, to name a few) on the numerical precision of the resulting $U$ and $V$ values to develop predefined sets of input parameters with a predictable output precision.
The present framework includes three such \texttt{protocols} named ``fast'', ``moderate'' and ``precise'', with their parameterization informed by several prior works by the authors.
It is generally advisable to include a geometry optimization step to avoid calculating Hubbard parameters for potentially unexpected ground states and to ensure consistency between the Hubbard parameters and the structure.
Specifically, while the differences in $U$ values between the workflows with and without structural optimization were minimal for most cases, a notable exception was observed for \ce{As2Fe2Li2}. 
In this case, the final $U_\mathrm{sc}$ parameter differed by over 1~eV due to a significant volume expansion during structural optimization.

Beyond such procedural details, a more fundamental aspect that is crucial for the use of Hubbard-corrected DFT methods is the choice of the Hubbard manifold.
To date and to the best of our knowledge, no unequivocal prescription has been developed that allows for a rational determination of the latter, i.e., to answer the question \textit{where to apply the $U$ (and the $V$, $J$ etc.)}.
The definition of the Hubbard manifold can influence the prediction of material properties~\cite{KirchnerHall2021, Moore2024, Macke2024}. 
For example, results may vary depending on whether $U$ corrections are applied to ligand $p$ shells or if the $4f$ shell, the $5d$ shell, or both are targeted in lanthanides.
In certain cases, the traditional practice of (exclusively) correcting TM $d$ shells can  result in diverging or oscillating Hubbard parameters, as observed in two of the unsuccessful workchains (see Supplementary Discussion).
Therefore, until a well-defined prescription is developed, based, e.g., on correcting self-interaction errors on the most localized representations, automated DFT$+U(+V)$ workflows -- including the present one -- will require human choices for the Hubbard manifolds to linearize the DFT functional.
Nonetheless, \texttt{aiida-hubbard} can facilitate the exploration of various Hubbard manifolds in a fast and reproducible way, helping to address this challenge in the future.

Finally, we expect our work to be particularly useful for modeling large systems (e.g., with defects) or for calculating observables whose evaluation potentially requires numerous independent calculations, such as vibrational properties.
Both types of applications benefit from the enhanced electronic structure description provided by DFT$+U(+V)$, which incurs minimal additional computational cost compared to (semi)local functionals.
This prospect is supported by the extendable and modular nature of the package, and even more so by the constantly growing universe of AiiDA plugins such as \texttt{aiida-vibroscopy}~\cite{Bastonero2024} and \texttt{aiida-defects}~\cite{Muy2023}.
%

\section*{METHODS}

\subsection*{Self-consistent calculation of \texorpdfstring{$U$}{U} and \texorpdfstring{$V$}{V}}
All calculations are performed using the \texttt{aiida-hubbard} plugin v.0.1.0 that is run using \QE distribution v7.2~\cite{Giannozzi2009, Giannozzi2017, Giannozzi2020, Timrov2022}.
We use the PBEsol flavor~\cite{Perdew2008} for the spin-polarized GGA xc functional, and employ pseudopotentials from the SSSP library v1.3 (efficiency)~\cite{Prandini2018}.
To construct the Hubbard projector functions we use atomic orbitals which are orthogonalized using the L\"owdin's method~\cite{Loewdin1950, Mayer2002, Timrov2020a}.
The Brillouin zone is sampled using uniform $\Gamma$-centered $\mathbf{k}$-point meshes with $\Delta \mathbf{k} = 0.2\,$\AA$^{-1}$.
The kinetic-energy cutoffs for the expansion of KS wavefunctions are set to the recommended values of the SSSP library~\cite{Prandini2018,Bosoni2023}. 
The crystal structures are optimized using the Broyden-Fletcher-Goldfarb-Shanno (BFGS) algorithm, with a convergence threshold for the total energy of $10^{-6}\,$Ry/atom, for forces of $10^{-4}\,$Ry/Bohr, and for pressure of $0.5\,$kbar.
For the DFT step prior to each DFPT calculation, we use Marzari-Vanderbilt cold smearing~\cite{Marzari1999} with a broadening parameter of 0.01\,Ry and $\Delta \mathbf{k} = 0.4\,$\AA$^{-1}$.
The DFPT calculations of Hubbard parameters are performed using $\mathbf{q}$-point meshes with a maximum distance of $\Delta \mathbf{q} = 0.8\,$\AA$^{-1}$.
As described above, structural optimizations, single-point DFT+$U$+$V$ calculations, and DFPT steps are iterated until self-consistency, which is achieved when the variation of Hubbard parameters fells below $\delta U = \delta V = 0.1$~eV.
At each iteration, only intersite parameters of the full $V^{IJ}$ belonging to the nearest neighbours of the TM ions are kept for each next iteration.
Nearest neighbour analysis is performed using the \texttt{CrystalNN}~\cite{Pan2021} class as implemented in Pymatgen~\cite{Ong2013, Isayev2017}, which exploits a Voronoi algorithm~\cite{Blatov*2004} to determine the number of nearest neighbours.
%


\section*{DATA AVAILABILITY}

The data used to produce the results of this work are available in the Materials Cloud Archive~\cite{Bastonero2025}.


\section*{CODE AVAILABILITY}

The code is open source and made available on GitHub (\href{https://github.com/aiidateam/aiida-hubbard}{https://github.com/aiidateam/aiida-hubbard}).
It is also distributed as an installable package through the Python Package Index (\href{https://pypi.org/project/aiida-hubbard/}{https://pypi.org/project/aiida-hubbard/}).
The base code is open to external contributions for improvements through the GitHub pull request system.
The full documentation with tutorials can be found at \href{https://aiida-hubbard.readthedocs.io/en/latest/}{https://aiida-hubbard.readthedocs.io/en/latest/}.
%


\section*{ACKNOWLEDGEMENTS}
L.B., C.M. and N.M. gratefully acknowledge support from the Deutsche Forschungsgemeinschaft (DFG) under Germany’s Excellence Strategy (EXC 2077, No. 390741603, University Allowance, University of Bremen) and Lucio Colombi Ciacchi, the host of the “U Bremen Excellence Chair Program”.
C.M. and E.M  acknowledge funding by MaX "Materials Design at the Exascale”, a Center of Excellence co-funded by the European High Performance Computing Joint Undertaking (JU) and participating countries under grant agreement No. 101093374.
M.B. acknowledges funding by the European Centre of Excellence MaX “Materials design at the Exascale” (grant no. 824143), and by the SwissTwins project, funded by the Swiss State Secretariat for Education, Research and Innovation (SERI).
We acknowledge support by the NCCR MARVEL, a National Centre of Competence in Research, funded by the Swiss National Science Foundation (Grant number 205602). This work was supported by a grant from the Swiss National Supercomputing Centre (CSCS) under project ID~465000416 (LUMI-G).
We thank Julian Geiger, Gabriel Joalland, Austin Zadoks and Timo Reents for useful discussions and feedbacks.


\section*{Author contributions statement}

We use in the following the CRediT (Contributor Roles Taxonomy) author statement.
L.B.: conceptualization, methodology, software, validation, formal analysis, data curation, writing -- original draft, visualization;
C.M.: methodology, validation, formal analysis, data curation;
E.M.: formal analysis, writing -- original draft;
M.B.: software;
S.P.H.: supervision, conceptualization, methodology, software;
I.T.: supervision, methodology, resources, project administration;
N.M.: supervision, methodology, project administration, funding acquisition.
All authors: writing -- review \& editing.

\section*{Additional information}

Correspondence should be addressed to L.B. 



\clearpage

\setcounter{figure}{0}
\renewcommand{\figurename}{Supplementary Figure}
\renewcommand{\thefigure}{\arabic{figure}}

\setcounter{table}{0}
\renewcommand{\tablename}{Supplementary Table}

\begin{titlepage}
\title{
    \large{
        Supplementary Information for \\
        ``First-principles Hubbard parameters with automated and reproducible workflows''
    }
}

\author{Lorenzo Bastonero}
\email{lbastone@uni-bremen.de}
\affiliation{U Bremen Excellence Chair, Bremen Center for Computational Materials Science, and MAPEX Center for Materials and Processes, University of Bremen, D-28359 Bremen, Germany}
% ...
\author{Cristiano Malica}
\affiliation{U Bremen Excellence Chair, Bremen Center for Computational Materials Science, and MAPEX Center for Materials and Processes, University of Bremen, D-28359 Bremen, Germany}
% ...
\author{Eric Macke}
\affiliation{U Bremen Excellence Chair, Bremen Center for Computational Materials Science, and MAPEX Center for Materials and Processes, University of Bremen, D-28359 Bremen, Germany}
% ...
\author{Marnik Bercx}
\affiliation{PSI Center for Scientific Computing,
Theory, and Data, and National Centre for Computational Design and Discovery of Novel Materials (MARVEL), 5232 Villigen PSI, Switzerland}
% ...
\author{Sebastiaan P. Huber}
\affiliation{Theory and Simulation of Materials (THEOS), and National Centre for Computational Design and Discovery of Novel Materials (MARVEL), \'Ecole Polytechnique F\'ed\'erale de Lausanne (EPFL), CH-1015 Lausanne, Switzerland}
% ...
\author{Iurii Timrov}
\affiliation{PSI Center for Scientific Computing,
Theory, and Data, and National Centre for Computational Design and Discovery of Novel Materials (MARVEL), 5232 Villigen PSI, Switzerland}
% ...
\author{Nicola Marzari}
\affiliation{U Bremen Excellence Chair, Bremen Center for Computational Materials Science, and MAPEX Center for Materials and Processes, University of Bremen, D-28359 Bremen, Germany}
\affiliation{PSI Center for Scientific Computing,
Theory, and Data, and National Centre for Computational Design and Discovery of Novel Materials (MARVEL), 5232 Villigen PSI, Switzerland}
\affiliation{Theory and Simulation of Materials (THEOS), and National Centre for Computational Design and Discovery of Novel Materials (MARVEL), \'Ecole Polytechnique F\'ed\'erale de Lausanne (EPFL), CH-1015 Lausanne, Switzerland}

\maketitle
\end{titlepage}

\clearpage

\section{Supplementary Discussion}
%
\subsection*{Analysis of the failed workchains}
Here we analyze the ten unsuccessful workchains that resulted from the investigation of the 115 Li-containing bulk structures.
An overview of the materials for which the workchain failed, along with additional information on the step of the workflow where it stopped and the reason of the failure, is provided in Supplementary Table \ref{tab:failed_wc}.
Two main causes of workflow failure were observed: (i) numerical issues, which account for eight of the ten cases, and (ii) physical issues related to the DFPT approach, causing divergence or oscillation of the computed Hubbard parameters in two cases.
At least some of the numerical issues can likely be resolved by further improving the determination and adaption of the input parameters so as to converge even the most difficult ground states.
Conversely, to understand the problem behind the workflows that did not converge due oscillating or diverging Hubbard $U$ parameters is less intuitive.
Specifically, the ``problematic'' compounds are the cubic Prussian Blue Analogues \ce{Li2Cu[Fe(CN)6]} and \ce{Cs2Li[Mn(CN)6]} (Hill formulae \ce{C6CuFeLi2N6} and \ce{C6Cs2LiMnN6}, respectively).
The occurrence of unphysical Hubbard $U$ values such as the $U\approx50\,eV$ observed for \ce{Cs2Li[Mn(CN)6]} can be understood from Eqs.~(2) and (5) of the main text: a Hubbard $U^I$ parameter diverges when $\chi^{II} \rightarrow 0$, i.e., when the perturbation $\alpha^J$ does not induce any measurable change in the occupations $\mathbf{n}^{II}$ of the Hubbard manifold.
While it has been recognized quite early that this can be an issue when computing Hubbard $U$ parameters of closed shells~\cite{Kulik2010, Yu2014} (e.g., the ${3d^{10}}$ shells of \ce{Zn^{2+}} or \ce{Cu^{1+}}), only recently attention has been brought to cases where $U$ diverges although the $d$ shell is incomplete~\cite{Macke2024}.
In fact, a vanishing $\chi$ is generally expected whenever the frontier states (HOMO and LUMO) contain no relevant contributions from the Hubbard manifold, e.g., from the TM-\textit{d} shell.
This often applies to insulators of the charge-transfer type and also to compounds with TM ions in low-spin configurations, as is the case for the \ce{Mn^3+} and \ce{Fe^3+} species in \ce{Cs2Li[Mn(CN)6]} and \ce{Li2Cu[Fe(CN)6]}, respectively~\cite{Mariano2021, Macke2024}.
Therefore, the failure of these two workchains has physical reasons rather than computational ones, and represents an irrecoverable error.
To resolve it, a revised and more suitable definition of the Hubbard manifold is required~\cite{Macke2024}.

\clearpage
\section{Supplementary Figures}

%
\begin{figure}[h!]
    \includegraphics[width=0.95\textwidth]{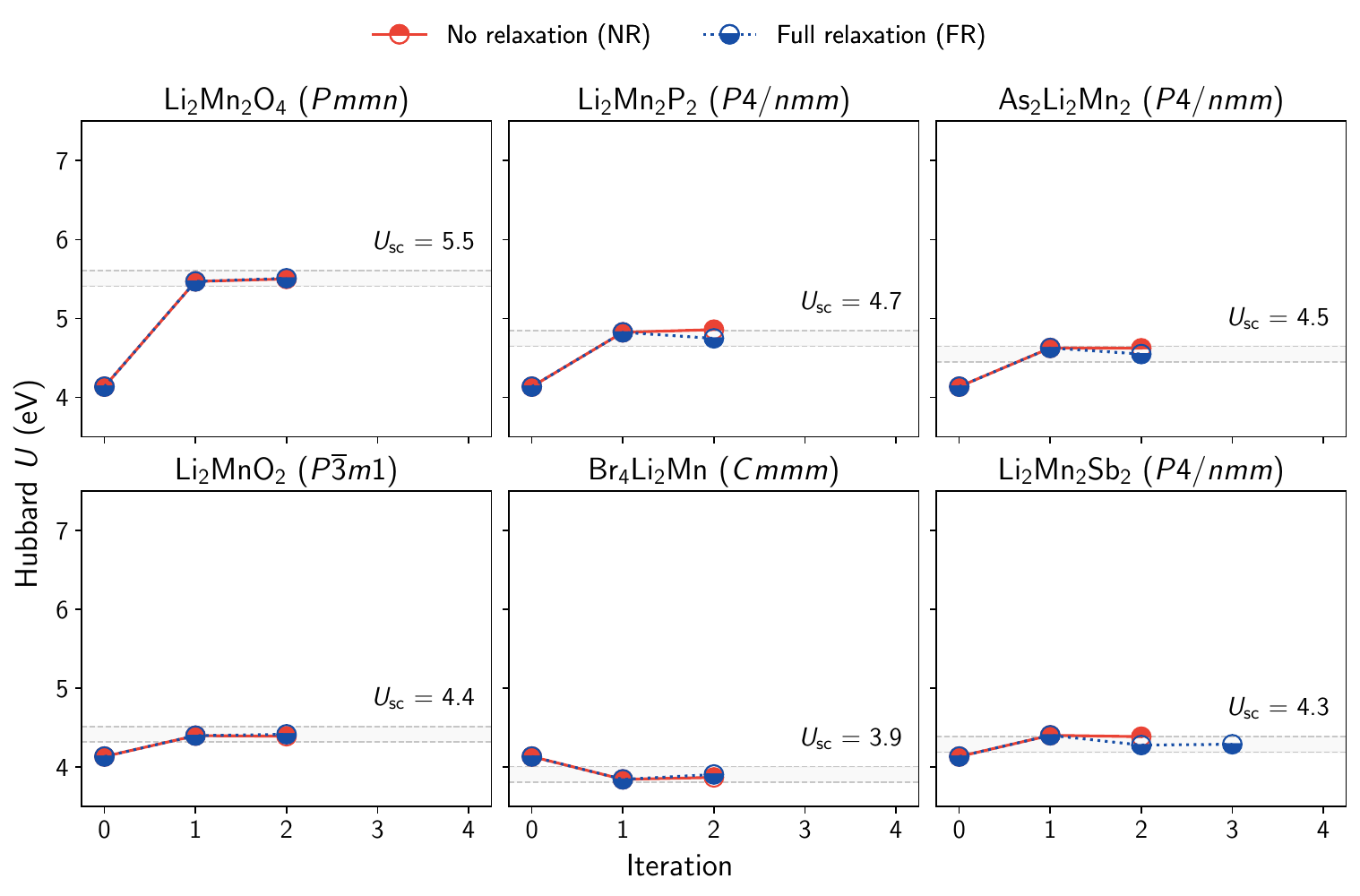}
    \caption{
        \textbf{Self-consistent convergence of Hubbard $U$ for Mn-$3d$ using two different strategies.} Values of Hubbard $U$ for six different Mn-containing bulk solids as a function of the iteration of the self-consistent cycle, using the NR (red) and FR (blue) schemes. For each compound, the associated Hill formula and space group is reported in the title of the corresponding panel. Iteration 0 corresponds to the starting guess $U_{\text{Mn}} = 4$~eV. Geometry optimizations of the FR cycle were omitted in iteration 1. The gray shaded area shows the convergence range $U_{\text{sc}} \pm \delta U$ ($\delta U = 0.1$~eV), around the final values of the FR cycles. 
    }
    \label{fig:sc_mn}
\end{figure}
%

\clearpage
\section{Supplementary Tables}
\begin{table}[h!]
    \centering 
    \caption{
        \textbf{Summary of the protocol main calculation parameters pre-defined by using the \texttt{get\_builder\_from\_protocol}.}
        List of parameters defined for the devised protocols available in the \texttt{aiida-hubbard} plugin v0.1.0. 
        Each protocol is identified by a string (``Name''), and it is associated to some pre-definition of calculation and computational resources parameters. 
        Here we report, in order from left to right: the pseudopotentials (kinetic and charge density cutoffs are set according to the associated cutoffs tables) taken from the corresponding SSSP library v1.3 for the PBEsol functional, the \textbf{k}-point distance for the DFT$+U+V$ and structural optimization step, the \textbf{q}-point distance for the DFTP calculation, and finally the thresholds for the Hubbard parameters $U$ and $V$ determining when the workflow reached self-consistency. 
        Other specific parameters can be found on the corresponding GitHub repositories, by using the Python API of the associated WorkChains, or by inspecting the available online documentation.
    }
    \label{tab:protocols}
    \begin{tabular}{lcccccc}
        \toprule
        Name & SSSP/1.3/PBEsol & $\Delta \mathbf{k}^{\mathrm{DFT}+U+V}$ & $\Delta \mathbf{k}^{\mathrm{Relax}}$ & $\Delta \mathbf{q}^{\mathrm{DFPT}}$ & $\delta U$ & $\delta V$ \\
        \midrule
        fast & efficiency & 0.60 & 0.50 & 1.20 & 0.20 & 0.10 \\
        moderate & efficiency & 0.40 & 0.15 & 0.80 & 0.10 & 0.01 \\
        precise & precision & 0.20 & 0.10 & 0.40 & 0.01 & 0.005 \\
        \bottomrule
    \end{tabular}
\end{table}
%
\LTcapwidth=\textwidth

\begin{longtable}{llccccccccr}
\caption{
\textbf{Summary of relevant data for the 105 Li-bearing materials.}
List of the 105 crystal structures for which the \texttt{Self\-Consistent\-Hubbard\-WorkChain} calculation completed successfully. The table reports the formula Hill of the material (``Formula''), its space group  (``SG''), the electronic band gap ($E_{\mathrm{gap}}$) computed using the last called \texttt{PwBaseWorkChain} calculation of the workflow (0.0 means metallic), the transition metal(s) (``TM'') and its oxidation state(s) and computed $U_{\mathrm{sc}}$, the range of computed $V_{\mathrm{sc}}$ and associated interatomic distances $r$, the number of $V_{\mathrm{sc}}$ interactions defined for the TM atom (N$_{V_{\text{sc}}}$), the number of atoms inside the unit cell, the first eight digits of the UUID (indicated by UUID$^*$) of the final \texttt{HubbardStructureData} associated with the crystal structure. Electronic band gap and Hubbard parameters are given in eV units, interatomic distances are in \AA~units. The list is ordered by TM first, then by its OS (ascending order), and finally by the numerical value of the associated $U_{\mathrm{sc}}$ (ascending order).
} \\
\toprule
Formula & SG & $E_{\mathrm{gap}}$ & TM & OS & $U_{\text{sc}}$ & $V_{\text{sc}}$ (min-max) & $r$ (min-max) & N$_{V_{\text{sc}}}$ & N$_{\text{at}}$ & UUID$^*$ \\
\midrule
\endfirsthead
\toprule
Formula & SG & $E_{\mathrm{gap}}$ & TM & OS & $U_{\text{sc}}$ & $V_{\text{sc}}$ (min-max) & $r$ (min-max) & N$_{V_{\text{sc}}}$ & N$_{\text{at}}$ & UUID$^*$ \\
\midrule
\endhead
\midrule
\multicolumn{11}{r}{Continued on next page} \\
\midrule
\endfoot
\bottomrule
\endlastfoot
$\ce{Fe4Li4O16P4}$ & $Pnma$ & 4.2 & Fe & 2 & 5.23 & 0.39 -- 0.88 & 2.06 -- 2.25 & 6 & 28 & 016fa32c \\
$\ce{As2Fe2Li2}$ & $P4/nmm$ & 0.0 & Fe & 2 & 5.37 & -- & -- & -- & 6 & 13a84d6c \\
$\ce{Ba4Fe4Li2N6}$ & $C2/c$ & 0.0 & Fe & 2 & 5.37 & -- & -- & -- & 16 & 1e0f3e74 \\
$\ce{Fe2Li4O8Si2}$ & $Pc$ & 3.4 & Fe & 2 & 5.38 & 0.5 -- 0.64 & 1.99 -- 2.06 & 4 & 16 & a3142187 \\
$\ce{Fe2Li2P2}$ & $Cmcm$ & 0.0 & Fe & 2 & 5.38 & -- & -- & -- & 6 & b6241509 \\
$\ce{Fe2Li2P2}$ & $P4/nmm$ & 0.0 & Fe & 2 & 5.40 & -- & -- & -- & 6 & 9e86853f \\
$\ce{B4Fe4Li4O12}$ & $C2/c$ & 3.5 & Fe & 2 & 5.40 & 0.26 -- 0.84 & 1.97 -- 2.29 & 5 & 24 & 83466658 \\
$\ce{Fe2Li4O8Si2}$ & $Pmn2_1$ & 3.5 & Fe & 2 & 5.40 & 0.51 -- 0.67 & 2.0 -- 2.07 & 4 & 16 & dd95032a \\
$\ce{Fe4Li8O16Si4}$ & $Pnma$ & 3.5 & Fe & 2 & 5.43 & 0.48 -- 0.7 & 1.99 -- 2.12 & 4 & 32 & e591a01f \\
$\ce{Fe4Li8O16Si4}$ & $P2_1/c$ & 3.5 & Fe & 2 & 5.43 & 0.51 -- 0.69 & 1.99 -- 2.09 & 4 & 32 & 467dde85 \\
$\ce{F2Fe2Li4O8P2}$ & $P1$ & 4.2 & Fe & 2 & 5.43 & 0.75 -- 0.99 & 2.05 -- 2.11 & 4 & 18 & fed35928 \\
$\ce{Fe2Li2O8P2}$ & $Cmcm$ & 3.9 & Fe & 2 & 5.45 & 0.54 -- 0.95 & 2.06 -- 2.19 & 6 & 14 & f01cfc97 \\
$\ce{As4Fe4Li4O16}$ & $Pnma$ & 2.3 & Fe & 2 & 5.46 & 0.42 -- 0.9 & 2.07 -- 2.27 & 6 & 28 & 5e922c44 \\
$\ce{Fe2Ge2Li4O8}$ & $Pmn2_1$ & 2.6 & Fe & 2 & 5.47 & 0.57 -- 0.74 & 2.0 -- 2.06 & 4 & 16 & 92020bef \\
$\ce{Fe4Ge4Li8O16}$ & $Pnma$ & 2.6 & Fe & 2 & 5.49 & 0.56 -- 0.77 & 1.99 -- 2.06 & 4 & 32 & 99dc4c84 \\
$\ce{Cl2Fe2Li2O8W2}$ & $P2_1/m$ & 3.5 & Fe & 2 & 5.97 & 0.72 -- 1.24 & 2.03 -- 2.29 & 4 & 16 & 6277cc5e \\
$\ce{Fe2Li2O8P2}$ & $P\overline{1}$ & 3.3 & Fe & 2 & 6.00 & 0.67 -- 1.01 & 1.95 -- 2.08 & 4 & 14 & e97dc3bb \\
$\ce{Cl2Fe2Li2Mo2O8}$ & $P2_1/m$ & 2.9 & Fe & 2 & 6.07 & 0.77 -- 1.3 & 2.03 -- 2.28 & 4 & 16 & 3b438081 \\
$\ce{Fe6Ge4Li}$ & $R\overline{3}m$ & 0.0 & Fe & 2 & 6.10 & -- & -- & -- & 11 & aafb22c5 \\
$\ce{FeLi2O8W2}$ & $P\overline{1}$ & 2.9 & Fe & 2 & 6.50 & 0.78 -- 1.15 & 2.05 -- 2.22 & 6 & 13 & 6e4b5fe5 \\
$\ce{F6FeLiRb2}$ & $Fm\overline{3}m$ & 3.3 & Fe & 3 & 4.43 & -- & -- & -- & 10 & 824cae1e \\
$\ce{F12Fe2Li2Rb4}$ & $R\overline{3}m$ & 3.1 & Fe & 3 & 4.45 & -- & -- & -- & 20 & 1b1931aa \\
$\ce{Ca2F12Fe2Li2}$ & $P\overline{3}1c$ & 3.6 & Fe & 3 & 4.51 & -- & -- & -- & 18 & 09b6f4a3 \\
$\ce{F18Fe3Li3Mn3}$ & $P321$ & 1.9 & \makecell{Fe\\Mn} & \makecell{3\\2} & \makecell{4.54\\3.60} & -- & -- & -- & 27 & d10e99d6 \\
$\ce{Cs4F12Fe2KLi}$ & $R\overline{3}m$ & 3.6 & Fe & 3 & 4.59 & -- & -- & -- & 20 & 9592f972 \\
$\ce{Cd2F12Fe2Li2}$ & $P\overline{3}1c$ & 3.2 & Fe & 3 & 4.59 & -- & -- & -- & 18 & 41628b5a \\
$\ce{F12Fe4Li2}$ & $P4_2nm$ & 1.7 & Fe & 3 & 4.62 & -- & -- & -- & 18 & 0f074e2f \\
$\ce{Br16Fe4Li4}$ & $P2_1/c$ & 1.6 & Fe & 3 & 4.65 & -- & -- & -- & 24 & b2a84fa1 \\
$\ce{Cl16Fe4Li4}$ & $P2_1/c$ & 2.2 & Fe & 3 & 4.65 & -- & -- & -- & 24 & d4348d92 \\
$\ce{Fe4Li4O8}$ & $Pna2_1$ & 1.7 & Fe & 3 & 4.88 & 0.74 -- 0.77 & 1.9 -- 1.91 & 4 & 16 & 38c1c476 \\
$\ce{Fe4Li4O16Si4}$ & $Pna2_1$ & 2.9 & Fe & 3 & 4.91 & 0.8 -- 0.85 & 1.88 -- 1.9 & 4 & 28 & 44219f86 \\
$\ce{Fe2Li2O8Si2}$ & $Pc$ & 2.9 & Fe & 3 & 4.94 & 0.76 -- 0.84 & 1.88 -- 1.91 & 4 & 14 & d8392931 \\
$\ce{FeLiO2}$ & $R\overline{3}m$ & 1.7 & Fe & 3 & 4.97 & 0.53 -- 0.53 & 2.04 -- 2.04 & 6 & 4 & 565f1e2b \\
$\ce{FeLiP}$ & $I4mm$ & 0.0 & Fe & 3 & 5.00 & -- & -- & -- & 3 & d13a74d8 \\
$\ce{AsFeLi}$ & $I4mm$ & 0.0 & Fe & 3 & 5.02 & -- & -- & -- & 3 & 53d2fb1a \\
$\ce{As2Fe2Li2}$ & $P6_3/mmc$ & 0.0 & Fe & 3 & 5.05 & -- & -- & -- & 6 & 2d5940ab \\
$\ce{Fe2Li2O4}$ & $I4_1/amd$ & 1.0 & Fe & 3 & 5.09 & 0.63 -- 0.66 & 2.02 -- 2.05 & 6 & 8 & ba2f2c65 \\
$\ce{C6Cs2FeLiN6}$ & $Fm\overline{3}m$ & 1.6 & Fe & 3 & 5.09 & -- & -- & -- & 16 & f3174ac0 \\
$\ce{Fe2Li2O12Si4}$ & $C2/c$ & 2.9 & Fe & 3 & 5.10 & 0.57 -- 0.97 & 1.92 -- 2.17 & 6 & 20 & c68d530e \\
$\ce{Fe2H2Li2O10P2}$ & $P\overline{1}$ & 2.4 & Fe & 3 & 5.22 & 0.71 -- 0.99 & 1.96 -- 2.06 & 6 & 18 & 58db6c75 \\
$\ce{Fe2Ge4Li2O12}$ & $C2/c$ & 2.5 & Fe & 3 & 5.22 & 0.68 -- 0.95 & 1.94 -- 2.13 & 6 & 20 & 7fc2690b \\
$\ce{C12Fe2Li2N12Rb4}$ & $P2_1/c$ & 1.9 & Fe & 3 & 5.22 & -- & -- & -- & 32 & b80935ff \\
$\ce{Fe2Li2O14P4}$ & $P2_1$ & 3.0 & Fe & 3 & 5.41 & 0.86 -- 1.08 & 1.95 -- 2.08 & 6 & 22 & 751f1d3b \\
$\ce{As2FeLiO7}$ & $C2$ & 2.6 & Fe & 3 & 5.54 & 0.88 -- 1.24 & 1.94 -- 2.05 & 6 & 11 & 9e4f1352 \\
$\ce{Fe2Li6N4}$ & $Ibam$ & 0.0 & Fe & 3 & 5.56 & -- & -- & -- & 12 & afa34d07 \\
$\ce{Fe2Li2O20Se8}$ & $Pnc2$ & 2.1 & Fe & 3 & 5.59 & 0.82 -- 1.17 & 2.0 -- 2.07 & 6 & 32 & 19fd0c67 \\
$\ce{FeLi4N2}$ & $Immm$ & 0.8 & Fe & 3 & 5.65 & -- & -- & -- & 7 & 9c509aa5 \\
$\ce{Fe2Li2Mo4O16}$ & $P\overline{1}$ & 2.3 & Fe & 3 & 5.73 & 0.93 -- 1.31 & 1.96 -- 2.06 & 6 & 24 & 6999e744 \\
$\ce{Fe2Li2O16W4}$ & $C2/c$ & 2.5 & Fe & 3 & 5.73 & 1.03 -- 1.13 & 1.98 -- 2.09 & 6 & 24 & b9a97395 \\
$\ce{Br4FeLi2}$ & $Cmmm$ & 0.0 & Fe & 3 & 5.75 & -- & -- & -- & 7 & 998cc0b5 \\
$\ce{Fe2Li4O16P4}$ & $P2_1/c$ & 0.2 & Fe & 3 & 5.86 & 1.07 -- 1.37 & 1.97 -- 2.09 & 6 & 26 & c5c3fa90 \\
$\ce{FeLi2S2}$ & $P\overline{3}m1$ & 0.0 & Fe & 3 & 5.93 & 0.34 -- 0.34 & 2.57 -- 2.57 & 6 & 5 & 0d7aaecb \\
$\ce{Fe2Li4S8Sn2}$ & $Pc$ & 0.0 & Fe & 3 & 5.93 & 0.47 -- 0.64 & 2.37 -- 2.39 & 4 & 16 & 7e48fbf6 \\
$\ce{Cl8FeLi6}$ & $Fm\overline{3}m$ & 0.0 & Fe & 3 & 6.21 & -- & -- & -- & 15 & c5abf8a0 \\
$\ce{FeLa2LiO6}$ & $R\overline{3}$ & 0.9 & Fe & 5 & 7.12 & 0.59 -- 0.59 & 1.86 -- 1.86 & 6 & 10 & 62db0762 \\
$\ce{F8HfLi2Mn}$ & $I\overline{4}$ & 5.7 & Mn & 2 & 3.58 & -- & -- & -- & 12 & af8b1875 \\
$\ce{F8Li2MnZr}$ & $I\overline{4}$ & 5.1 & Mn & 2 & 3.73 & -- & -- & -- & 12 & 24fece8d \\
$\ce{F18Li3Mn3Ti3}$ & $P321$ & 4.1 & \makecell{Mn\\Ti} & \makecell{2\\3} & \makecell{3.75\\5.16} & -- & -- & -- & 27 & 88c37e8a \\
$\ce{Br4Li2Mn}$ & $Cmmm$ & 3.5 & Mn & 2 & 3.91 & -- & -- & -- & 7 & 9004cf7d \\
$\ce{F18Li3Mn3V3}$ & $P321$ & 2.8 & \makecell{Mn\\V} & \makecell{2\\3} & \makecell{3.91\\4.47} & -- & -- & -- & 27 & 64e44887 \\
$\ce{Li8Mn2Se14Sn4}$ & $Cc$ & 1.5 & Mn & 2 & 4.19 & 0.12 -- 0.13 & 2.57 -- 2.59 & 4 & 28 & 0a8bc9a6 \\
$\ce{LiMnTe2}$ & $P3m1$ & 0.0 & Mn & 2 & 4.29 & 0.17 -- 0.22 & 2.67 -- 2.75 & 4 & 4 & 78fac50c \\
$\ce{Li2Mn2Sb2}$ & $P4/nmm$ & 0.0 & Mn & 2 & 4.31 & -- & -- & -- & 6 & dff4cc5c \\
$\ce{Li4Mn4O16P4}$ & $Pnma$ & 3.2 & Mn & 2 & 4.36 & 0.38 -- 0.47 & 2.18 -- 2.24 & 6 & 28 & 184981a5 \\
$\ce{Li4Mn2S8Sn2}$ & $Pmn2_1$ & 1.9 & Mn & 2 & 4.38 & 0.11 -- 0.18 & 2.44 -- 2.46 & 4 & 16 & 1eeb5e7d \\
$\ce{Li8Mn4S16Sn4}$ & $Pna2_1$ & 2.0 & Mn & 2 & 4.43 & 0.17 -- 0.17 & 2.43 -- 2.45 & 4 & 32 & 1b301d2a \\
$\ce{Li4Mn2S8Sn2}$ & $Pc$ & 2.0 & Mn & 2 & 4.44 & 0.17 -- 0.18 & 2.43 -- 2.45 & 4 & 16 & a5514012 \\
$\ce{Ge4Li8Mn4S16}$ & $Pna2_1$ & 2.1 & Mn & 2 & 4.45 & 0.14 -- 0.16 & 2.42 -- 2.45 & 4 & 32 & fb561bdb \\
$\ce{Ge4Li8Mn2S14}$ & $Cc$ & 2.4 & Mn & 2 & 4.45 & 0.15 -- 0.19 & 2.42 -- 2.47 & 4 & 28 & c03f0795 \\
$\ce{Li2Mn2O8}$ & $I4_1/amd$ & 0.0 & Mn & 2 & 4.56 & 0.48 -- 0.84 & 2.15 -- 2.64 & 12 & 12 & ce87c8fe \\
$\ce{As2Li2Mn2}$ & $P4/nmm$ & 0.0 & Mn & 2 & 4.57 & -- & -- & -- & 6 & a085ae77 \\
$\ce{Li2MnO2}$ & $P\overline{3}m1$ & 2.6 & Mn & 2 & 4.57 & 0.22 -- 0.22 & 2.25 -- 2.25 & 6 & 5 & 14989b62 \\
$\ce{Li2Mn2Na2O8Si2}$ & $Pc$ & 2.7 & Mn & 2 & 4.61 & 0.54 -- 0.58 & 2.05 -- 2.07 & 4 & 16 & 78af6e7a \\
$\ce{AsLiMn}$ & $F\overline{4}3m$ & 0.1 & Mn & 2 & 4.71 & -- & -- & -- & 3 & 25503b48 \\
$\ce{Li2Mn2O8P2}$ & $Cmcm$ & 3.3 & Mn & 2 & 4.71 & 0.56 -- 0.8 & 2.13 -- 2.21 & 6 & 14 & 004e178f \\
$\ce{As4Li4Mn4O16}$ & $Pnma$ & 1.7 & Mn & 2 & 4.74 & 0.42 -- 0.81 & 2.13 -- 2.29 & 6 & 28 & 8ce9b10d \\
$\ce{Li2Mn2P2}$ & $P4/nmm$ & 0.0 & Mn & 2 & 4.74 & -- & -- & -- & 6 & d5c1a4b5 \\
$\ce{K2Li2Mn2O4}$ & $C2/m$ & 1.3 & Mn & 2 & 4.75 & 0.24 -- 0.46 & 2.04 -- 2.18 & 4 & 10 & 2c4f07e9 \\
$\ce{B3Li3Mn3O9}$ & $P\overline{6}$ & 2.7 & Mn & 2 & 4.81 & 0.44 -- 0.59 & 2.08 -- 2.17 & 5 & 18 & b74fe341 \\
$\ce{Li4Mn2O8Si2}$ & $Pmn2_1$ & 3.2 & Mn & 2 & 4.85 & 0.48 -- 0.54 & 2.06 -- 2.09 & 4 & 16 & 3cace8d4 \\
$\ce{Li8Mn4O16Si4}$ & $Pnma$ & 3.2 & Mn & 2 & 4.85 & 0.48 -- 0.57 & 2.06 -- 2.09 & 4 & 32 & 39ceaf22 \\
$\ce{Li4Mn2O8Si2}$ & $Pc$ & 3.1 & Mn & 2 & 4.86 & 0.47 -- 0.58 & 2.04 -- 2.08 & 4 & 16 & 9f68f1d1 \\
$\ce{Ge2Li4Mn2O8}$ & $Pmn2_1$ & 2.5 & Mn & 2 & 4.90 & 0.54 -- 0.6 & 2.06 -- 2.09 & 4 & 16 & d747d36d \\
$\ce{LiMnSe2}$ & $P3m1$ & 0.0 & Mn & 2 & 4.91 & 0.31 -- 0.4 & 2.46 -- 2.55 & 4 & 4 & 4ea95a4f \\
$\ce{Li2Mn2O8V2}$ & $Cmcm$ & 3.1 & \makecell{Mn\\V} & \makecell{2\\5} & \makecell{5.33\\5.85} & \makecell{0.64 -- 0.75\\1.25 -- 1.28} & \makecell{2.16 -- 2.2\\1.69 -- 1.75} & \makecell{6\\4} & 14 & 3a23d201 \\
$\ce{BaLi2MnO8V2}$ & $P\overline{3}$ & 3.8 & \makecell{Mn\\V} & \makecell{2\\5} & \makecell{5.31\\5.71} & \makecell{0.75 -- 0.75\\1.34 -- 1.38} & \makecell{2.18 -- 2.18\\1.70 -- 1.73} & \makecell{6\\4}  & 14 & be9b8a1d \\
$\ce{KLiMn2O12Si4}$ & $C2/m$ & 2.3 & Mn & 3 & 5.64 & 0.4 -- 0.78 & 1.88 -- 2.26 & 6 & 20 & 294b9fe4 \\
$\ce{F10Li4Mn2}$ & $C2/c$ & 2.1 & Mn & 3 & 5.66 & -- & -- & -- & 16 & c4a9c584 \\
$\ce{Li2Mn2O4}$ & $Pmmn$ & 1.4 & Mn & 3 & 5.75 & 0.29 -- 0.62 & 1.92 -- 2.31 & 6 & 8 & a2d005d1 \\
$\ce{F8Li2Mn2}$ & $P2_1/c$ & 2.2 & Mn & 3 & 5.79 & -- & -- & -- & 12 & abcf4222 \\
$\ce{Li4Mn4O8}$ & $I4_1/amd$ & 1.5 & Mn & 3 & 5.81 & 0.29 -- 0.63 & 1.94 -- 2.33 & 6 & 16 & eddfe602 \\
$\ce{F8Li6Mn2O12P4}$ & $P2_1/c$ & 2.0 & Mn & 3 & 6.08 & 0.7 -- 0.91 & 1.91 -- 2.19 & 4 & 32 & 1f4176f0 \\
$\ce{H2Li2Mn2O10P2}$ & $P\overline{1}$ & 1.2 & Mn & 3 & 6.29 & 0.64 -- 1.03 & 1.9 -- 2.25 & 6 & 18 & 787f2595 \\
$\ce{Li2Mn2O14P4}$ & $P2_1$ & 1.9 & Mn & 3 & 6.34 & 0.76 -- 1.05 & 1.91 -- 2.19 & 6 & 22 & 11150a88 \\
$\ce{Ca2Li6Mn2N6}$ & $R\overline{3}$ & 0.7 & Mn & 4 & 5.98 & -- & -- & -- & 16 & 5df0d47e \\
$\ce{Li8Mn4O12}$ & $C2/c$ & 2.1 & Mn & 4 & 6.49 & 0.68 -- 0.69 & 1.92 -- 1.93 & 6 & 24 & c78b6e9c \\
$\ce{Li4Mn2O6}$ & $C2/m$ & 1.9 & Mn & 4 & 6.49 & 0.68 -- 0.69 & 1.92 -- 1.93 & 6 & 12 & 6b53bc4b \\
$\ce{Li4Mn4Ni2O12}$ & $Cmce$ & 1.6 & \makecell{Mn\\Ni} & \makecell{4\\2} & \makecell{6.51\\5.16} & \makecell{0.6 -- 0.61\\0.38 -- 0.47} & \makecell{1.9 -- 1.95\\2.03 -- 2.1} & 6 & 22 & 504172a5 \\
$\ce{Li2Mn2O10P2}$ & $P\overline{1}$ & 0.0 & Mn & 4 & 7.23 & 0.94 -- 1.47 & 1.85 -- 2.11 & 6 & 16 & a44df345 \\
$\ce{Li2Mn4O8}$ & $Fd\overline{3}m$ & 0.0 & Mn & 4 & 9.21 & 1.29 -- 1.29 & 2.01 -- 2.01 & 6 & 14 & 76ad08cf \\
$\ce{K11LiMn4O16}$ & $I\overline{4}2m$ & 2.1 & Mn & 5 & 5.40 & 0.63 -- 0.65 & 1.7 -- 1.7 & 4 & 32 & 9eb5676a \\
$\ce{Li6Mn2O8}$ & $Pmn2_1$ & 2.1 & Mn & 5 & 5.80 & 0.68 -- 0.73 & 1.7 -- 1.71 & 4 & 16 & ad395cda \\
$\ce{Cs4Li2Mn2O8}$ & $Cmc2_1$ & 2.2 & Mn & 5 & 5.82 & 0.63 -- 0.73 & 1.69 -- 1.72 & 4 & 16 & 61002306 \\
$\ce{Li12Mn4O16}$ & $Pnma$ & 2.1 & Mn & 5 & 5.82 & 0.72 -- 0.73 & 1.7 -- 1.71 & 4 & 32 & d6d86f57 \\
$\ce{Li2Mn2O8}$ & $Cmcm$ & 1.7 & Mn & 7 & 6.55 & 0.5 -- 0.63 & 1.59 -- 1.61 & 4 & 12 & c745c192 \\
\label{tab:success_wc}
\end{longtable}

\newpage
\begin{longtable}{lllp{0.4\textwidth }}
\caption{
\textbf{Information about the 10 failed calculations.}
List of the 10 crystal structures for which the \texttt{Self\-Consistent\-Hubbard\-WorkChain} calculation stopped and the self-consistent Hubbard parameters have not been computed. The table reports the formula Hill of the material (``Formula''), its space group  (``SG''), the calculation step where workflow was interrupted (``Step of failure''), and the message describing the reason of the failure (``Exit message'').
} \\
\toprule
Formula & SG & Step of failure & Exit message \\
\midrule
\endfirsthead
\toprule
Formula & SG & Step of failure & Exit message \\
\midrule
\endhead
\midrule
\multicolumn{4}{r}{Continued on next page} \\
\midrule
\endfoot
\bottomrule
\endlastfoot
$\ce{As4Fe4Li4}$ & $P\overline{3}m1$ & DFT$+U+V$ & The electronic minimization cycle failed during an ionic minimization cycle. \\
$\ce{Fe2Li2O8}$ & $Fd\overline{3}m$ & DFT$+U+V$ & The S matrix was found to be not positive definite. \\
$\ce{Br8Li2Mn8}$ & $I4_1/amd$ & DFT$+U+V$ & The S matrix was found to be not positive definite. \\
$\ce{As2Li2Mn2O10}$ & $P\overline{1}$ & Structure Optimization & The electronic minimization cycle did not reach self-consistency. \\
$\ce{F12Li2Mn2V2}$ & $P4_2nm$ & Structure Optimization & The stdout output file was incomplete probably because the calculation got interrupted. \\
$\ce{F6FeLi3}$ & $Fm\overline{3}m$ & DFPT & The stdout output file was incomplete probably because the calculation got interrupted. \\
$\ce{Br8Li4Mn2}$ & $Imma$ & DFPT & The code failed due to incompatibility between the FFT grid and the parallelization options. \\
$\ce{Li4Mn2O8}$ & $Fddd$ & DFPT & The code failed due to incompatibility between the FFT grid and the parallelization options. \\
$\ce{C6Cs2LiMnN6}$ & $Fm\overline{3}m$ & Max iteration reached & -- \\
$\ce{C6CuFeLi2N6}$ & $Fm\overline{3}m$ & Max iteration reached & -- \\
\label{tab:failed_wc}
\end{longtable}



%


\clearpage

\bibliography{main}

\end{document}